\documentstyle[sprocl]{article}

\input epsf.sty
\input psfig.sty
\def\ga{\mathrel{\raise.3ex\hbox{$>$\kern-.75em\lower1ex\hbox{$\sim$}}}}
\def\la{\mathrel{\raise.3ex\hbox{$<$\kern-.75em\lower1ex\hbox{$\sim$}}}}
\def\he#1{\hbox{${}^{#1}$He}}
\def\li#1{\hbox{${}^{#1}$Li}}
\def\be#1{\hbox{${}^{#1}$Be}}
\def\b#1#2{\hbox{${}^{#1#2}$B}}
\def\beq{\begin{equation}}
\def\eeq{\end{equation}}

\begin{document}
\vspace*{-2.5cm}
\rightline{UMN--TH--1539/96}
\rightline{astro-ph/9707212}
\rightline{May 1997}

\title{Primordial Nucleosynthesis and Dark Matter$^a$}\footnotetext{
$^a$Summary of lectures given at the 1997 Lake Louise Winter Institute,
  Lake Louise, Alberta, February 16-20 1997.}

\author{Keith A.~Olive}

\address
{School of Physics and Astronomy,
University of Minnesota,\\ Minneapolis, MN 55455, USA}

\maketitle\abstracts{
Following a brief introduction to early Universe cosmology,
the current of status of big bang nucleosynthesis is reviewed and
the concordance between theory and observation is examined in
detail.  The abundances of \he4 and \li7  determine
the value of the baryon-to-photon ratio, $\eta$ to be relatively low,
$\eta \approx 1.8 \times 10^{-10}$, and agrees with 
some recent measurements of D/H in quasar absorption
systems.  These results have far reaching consequences for
galactic chemical evolution, the amount of baryonic dark
matter in the Universe and on the allowed number of degrees of freedom
in the early Universe. Finally, motivations for cosmological dark matter
will be reviewed with special emphasis placed on supersymmetric
candidates. }


\section{Introduction}
To set the framework for the subsequent discussion on 
Big Bang Nucleosynthesis (BBN) and dark matter, 
it will be useful to briefly review
some general aspects of the standard hot big bang model.   BBN occurs
at the relatively late time of order 1 minute after the big bang,
however, the initial conditions which determine BBN, were set up
much earlier.  These include, an isotropic background, a finite 
baryon density and a period of equilibrium.   
The standard big bang model assumes homogeneity and
isotropy, though it is reasoned that these conditions were
achieved after a period of inflation. Given homogeneity and isotropy,
 space-time can be described by the 
Friedmann-Robertson-Walker metric which in co-moving coordinates is given by
\beq
	ds^2  = -dt^2  + R^2(t)\left[ {dr^2 \over \left(1-kr^2\right) }
      + r^2 \left(d\theta^2  + \sin^2 \theta d\phi^2 \right)\right]	
\label{met}
\eeq
where $R(t)$ is the cosmological scale factor and $k$ is the three-space
curvature constant ($k = 0, +1, -1$ for a spatially flat, closed or open
Universe). $k$ and $R$ are the only two quantities in the
metric which distinguish it from flat Minkowski space.
It is  also common to assume
 the perfect fluid form for the energy-momentum
tensor
 described by, the isotropic
pressure, $p$, and the energy density $\rho$.
   Einstein's equations  yield the
Friedmann equation,
\beq
	H^2  \equiv \left({\dot{R} \over R}\right)^2  = {1 \over 3} 8 \pi G_N \rho
 - { k \over R^2}  + {1 \over 3} \Lambda
\label{H}
\eeq
and
\beq
	\left({\ddot{R} \over R}\right) = {1 \over 3} \Lambda -
 {1 \over 6} 8 \pi G_N ( \rho + 3p)
\eeq
where $\Lambda$ is the cosmological constant,
or equivalently from the conservation of the energy momentum tensor
\beq
	\dot{\rho} = -3H(\rho + p)	
\label{rhod}	
\eeq
These equations form the basis of the standard big bang model.

At early times ($t < 10^5 $ yrs) the Universe is thought to have been
dominated by radiation so that the equation of state can be given by $p =
\rho/3$.  Neglecting the contributions to $H$ from $k$ and $\Lambda$
 (this is always a
good approximation for small enough $R$), we find that
\beq
	R(t) \sim t^{1/2} \qquad    
\eeq
and $\rho \sim R^{-4}$  so that $t \sim (3/32 \pi G_N\rho)^{1/2}$.
  Similarly for a matter or dust
dominated Universe with $p = 0$,
\beq
	R(t) \sim t^{2/3}    
\eeq
and $\rho \sim R^{-3}$.  The Universe makes the transition
 between radiation and matter
domination when $\rho_{rad} = \rho_{matter}$ or 
when $T \simeq$ few $\times~10^3$ K.

In the absence of a cosmological constant, one can 
define a critical energy density $\rho_c$
  such that $\rho =\rho_c$  for $k = 0$
\beq
	\rho_c  = 3H^2 / 8 \pi G_N		
\eeq
In terms of the present value of the Hubble parameter this is,
\beq
	\rho_c  = 1.88 \times 10^{-29} {h_o}^2  {\rm g cm}^{-3}  		
\eeq
where
\beq
	h_o  = H_o /(100 {\rm km Mpc}^{-1}   {\rm s}^{-1}  )		
\eeq
The cosmological density parameter is then defined by
\beq
	\Omega \equiv {\rho \over \rho_c} 			
\eeq
in terms of which the Friedmann equation, Eq. (\ref{H}), can be rewritten as
\beq
	(\Omega - 1)H^2  = {k \over R^2}	
\label{o-1}	
\eeq
so that $k = 0, +1, -1$ corresponds to $\Omega = 1, \Omega > 1$
 and $\Omega < 1$.  
Broad observational limits on $h_o$ and $\Omega$ are\cite{tonry}
\beq
0.4 \le h_o \le 1.0 \qquad 0.1 \le \Omega \le 2
\label{range}
\eeq

The value of $\Omega$, at least on relatively small scales, seems to 
depend on scale. Indeed, the contribution to $\Omega$ from visible
matter associated with stars and hot gas is quite small, $\Omega \approx
0.003 - 0.01$.  On somewhat larger scales, that of galactic halos or
small groups of galaxies, $\Omega \approx 0.02 - 0.1$.
On galaxy cluster scales, it appears that $\Omega$ may be as large as 
0.3.  And while there is some evidence (see the lectures of 
N. Kaiser\cite{nkaiser} in these proceedings), the observations are
far from conclusive in indicating a value of $\Omega$ as large as 1.
It is however possible to obtain a bound on the product, $\Omega h^2$
from 
\beq
H_o t_o = \int_0^1 ( 1 - \Omega + \Omega/x )^{-1/2} dx
\eeq
(for $\Lambda = 0$).
For $t_o > 12$Gyr, and $\Omega \le 1$, $\Omega h^2 < 0.3$
(This is true even if $\Lambda \ne 0$.)

As indicated above,  BBN  takes place during the radiation dominated
epoch which lasts roughly to the period of
recombination (somewhat earlier when dark matter is included)
 which occurs when electrons
 and protons form neutral hydrogen  through 
$e^{ -  } +  p     \rightarrow $  H  $+   \gamma $
  at a temperature  
$T_{ R} { \sim }$  few $\times 10^{3}$ K  ${ \sim}1$ eV.  For $T < T_{R}$, 
photons are decoupled while for $T > T_{ R}$,  photons are
 in thermal equilibrium.  Today, the content
 of the microwave background consists of photons with
$T_o =  2.728  \pm .002$ K \cite{cobet}.  
The energy density of photons in the background can be calculated from
\beq
 \rho_\gamma  = \int E_\gamma dn_\gamma
\label{rhog}
\eeq
 where the density of states is given by
\beq
dn_\gamma  =   {g_\gamma \over 2 
 \pi^{ 2}}[exp(E_\gamma/T)-1]^{ -  1} q^{ 2} dq 
\eeq
 and $g_\gamma = $  2 is the number of spin polarizations for the 
photon,
$E_\gamma =  q$ is just the photon energy (momentum). 
 (I am using units such that  
$\hbar =  c  = k_{ B}   =$  1 and will do so through the remainder
 of these lectures.)  
Integrating (\ref{rhog}) gives
\beq
\rho_\gamma = {\pi^2 \over 15} T^4
\eeq
 which is the familiar blackbody result.

In general, at very early times, at very high temperatures,
 other particle degrees of freedom join the radiation background when  
$T{ \sim } m_{i}$  for each  particle type $i$ if that type is brought
 into thermal equilibrium through interactions.  In equilibrium 
the energy density of a particle type $i$ is given by
\beq
 \rho_{i}  = \int E_{i} dn_{q_{i}} 
\eeq
 and
\beq 
 dn_{q_{i}} = {g_{i} \over 2  \pi^{ 2}}[exp[(E_{q_{i}} - 
\mu_{i})/T] \pm 1]^{ -1 }q^{2}dq
\eeq
where again $g_{i}$ counts the total number of degrees of freedom for type i,
\beq
 E_{q_{i}} =  \left(m_{i}^{2} + q_{i}^{ 2}\right)^{1/2}
\eeq
$\mu_{i}$ is the chemical potential if present and  $ \pm$  
corresponds to either Fermi or Bose statistics.

In the limit that  $T \gg m_{i} $  the total energy density can
 be conveniently expressed by  
\beq
 \rho {} = \left( \sum_B g_{B} + {7 \over 8} \sum_F  g_{F} \right)
   {\pi^{ 2} \over 30}  T^{4}     \equiv    {\pi^{ 2} \over 30} N(T) T^{4} 
\label{NT}
\eeq
 where $g_{B(F)} $  are the total number of boson (fermion) 
degrees of freedom and the sum runs over all boson (fermion) states with 
$m \ll T$.  The factor of 7/8 is due to the difference between
 the Fermi and Bose integrals.  Equation (\ref{NT}) defines N(T)
 by taking into account  new particle degrees 
of freedom as the temperature is raised.  

In the radiation dominated epoch,
eq. (\ref{rhod}) can be integrated (neglecting the $T$-dependence of $N$)
giving us a relationship between
 the age of the Universe and its temperature
\beq
 t = \left({90 \over 32 \pi^3 G_{ N} N(T)}\right)^{ 1/2}  T^{ -  2} 
\label{tt1}
\eeq
 Put into a more convenient form
 \beq
 tT_{ MeV}^{ 2}  =
2.4 [N(T)]^{ -  1/2}  
\label{tt2}   
\eeq
 where t is measured in seconds and
$T_{ MeV} $  in units of MeV.

 The value of $N(T)$ at any given temperature depends
 on the particle physics model.  In the standard $SU(3) \times
SU(2)\times U(1)$  model, we can specify $N(T)$ up to 
temperatures of 0(100) GeV.
  The change in N can be seen in the following table.

\begin{table}
\caption{Effective numbers of degrees of freedom in the standard model.}
\vspace{2pt}
\begin{center}
\begin{tabular}{llc}
\hline
Temperature & New Particles \qquad
&$4N(T)$ \\
\hline\rule{0pt}{12pt}
$T < m_{ e}$   &     $\gamma$'s +   $\nu$'s & 29 \\
$m_{ e} <   T  < m_\mu$ &    $e^{\pm}$ & 43 \\
$m_\mu <  T  < m_\pi$  &   $\mu {}^{\pm}$ & 57 \\
$m_\pi <  T < T{ c}^{*}$  & $\pi$'s & 69 \\
$T_{ c} <  T  < m_{\rm strange~~~~~~~}$ \qquad &
  -  $\pi$'s + $  u,{\bar u},d,{\bar d}$ + gluons &  205 \\
$m_{ s} <  T < m_{ charm}$ & $s,{\bar s}$ & 247 \\
$m_{ c} <  T < m_\tau$ &  $c,{\bar c}$ & 289 \\
$m_\tau < T < m_{ bottom}$ & $\tau {}^{\pm}$ & 303 \\
$m_{ b} < T < m_{ W,Z}$ & $b,{\bar b}$ & 345 \\
$m_{ W,Z} <  T < m_{ top}$ & $W^{\pm}, Z$ & 381 \\
$ m_t < T < m_{Higgs}$ & $t,{\bar t}$ & 423 \\
$M_H < T $ & $H^o$ & 427 \\
\hline
\end{tabular}
\end{center}
*$T_{ c}$ 
corresponds to the confinement-deconfinement transition between
 quarks and hadrons.
 $N(T)$  is shown in Figure 1 for $T_{ c}  =  150$ and $400$ MeV.
It has been assumed that $m_{Higgs} > m_{top}$.
\end{table}

 At higher temperatures ($T \gg 100$ GeV), $N(T)$ will be model dependent.
  For example, in the minimal $SU(5)$ model, one needs to add
 to $N(T)$, 24 states for the X and Y gauge bosons, 
another 24 from the adjoint Higgs, and another 6 (in addition
to the 4 already counted in $W^\pm, Z$ and $H$) from the $ {\bar 5}$
of Higgs.
  Hence for   
$T > M_{ X}  $  in minimal $SU(5)$, $N(T)  =  160.75$. 
 In a supersymmetric model this would at least double, 
with some changes possibly necessary in the table if the
 lightest supersymmetric particle has a mass below  
$M_{ H}.$

\begin{figure}
\hspace{0.5truecm}
\epsfysize=7truecm\epsfbox{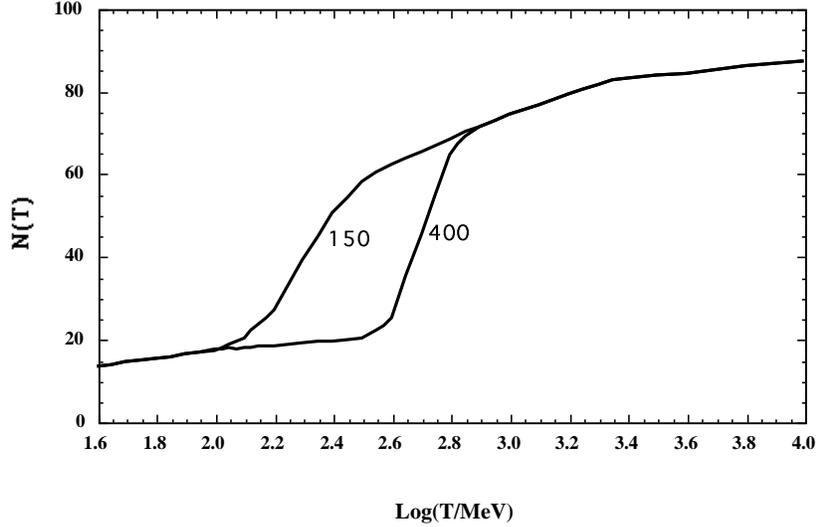}
\caption{The effective numbers of relativistic degrees of freedom
as a function of temperature.}
\end{figure}

The presence of a particle species in the thermal background 
assumes thermal equilibrium and hence interaction rates which
are sufficiently fast compared with the expansion rate of the 
Universe.
Roughly, this translates to the condition 
 for each particle type $i$, that some rate  
$  \Gamma {}_{ i} $  involving that type be larger than the expansion
 rate of the Universe or 
\beq
  \Gamma {}_{ i}  > H
\eeq
  in order to be in thermal equilibrium.

Examples of a  processes in equilibrium at early times 
 which drops out of equilibrium or decouples at later times
 are the processes which involve neutrinos.  
If we consider the standard neutral or charged-current interactions such as 
$e^{ +} + e^{ -  }     \leftrightarrow {}   \nu {}  +  \bar \nu $  
or $e +   \nu {}    \leftrightarrow {}  e  +   \nu {}$  etc.,
the rates for these processes can be approximated by
 \beq
 \Gamma {} =  n \langle \sigma v \rangle
 \eeq
 where  $  \langle \sigma v \rangle $  
is the thermally averaged  weak interaction cross section 
\beq
 \langle \sigma v \rangle { \sim }~0(10^{ -  2}) T^{ 2} /M_{ W}^4
\eeq
and $n$ is the number density of leptons.
 Hence the rate for these interactions is 
\beq
 \Gamma {}_{\rm wk}   { \sim  }~0(10^{ -  2}) T^{ 5}/M_{ W}^4
\eeq 
The expansion rate, on the other hand, is just
\beq
  H  =  \left({8 \pi {}G_{ N}  \rho {} \over 3}\right)^{ 1/2}   
 =  \left({8  \pi {}^{ 3} \over 90}  N(T) \right)^{ 1/2}  T^{ 2}/M_{ P}    
 \sim 1.66 N(T)^{ 1/2}  T^{ 2}/M_{ P}.  
\eeq
The Planck mass $M_{ P} = G_N^{-1/2} =  1.22 \times 10^{19}$ GeV.

Neutrinos will be in equilibrium when  $  \Gamma {}_{\rm wk} >  H $ or
\beq
T > (500 M_{ W}^4)/M_{ P})^{ 1/3}  { \sim }~1 MeV . 
\eeq
The temperature at which these rates are equal
 is commonly referred to as the decoupling or freeze-out 
temperature and is defined by 
\beq
  \Gamma(T_{ d}) = H(T_{ d}) 
\eeq
 For temperatures $T > T_{ d}$,  
neutrinos will be in equilibrium, while for $T < T_{ d }$ 
 they will not. Basically, in terms of their interactions, 
the expansion rate is just too fast and they never  
{\em ``see"}  the rest of the matter in the Universe (or themselves).
  Their momenta will simply redshift and their effective temperature 
(the shape of their momenta distribution is not changed from that 
of a blackbody) will simply fall with  
$T { \sim  } 1/R.$

 Soon after decoupling the $e^{\pm}$  pairs in the thermal background
begin to annihilate (when $T \la m_e$).  
Because the neutrinos are decoupled, 
the energy released heats up the photon background relative 
to the neutrinos. The change in the photon temperature can be
easily computed from entropy conservation. The neutrino entropy 
must be conserved separately from the entropy of interacting particles.
 If we denote $T_>$, the temperature of photons, and $e^{\pm}$
 before annihilation, we also have $T_\nu  = T_>$  as well. 
 The entropy density of the interacting particles at $T = T_>$ is just
\beq
	s_> =  {4 \over 3}  {\rho_> \over T_>}  = 
{4 \over 3} (2 + {7 \over 2}) ( {\pi^2 \over 30} ) T^3_>	
\eeq
while at $T = T_<$, 
the temperature of the photons just after $e^{\pm}$ annihilation, 
the entropy density is
\beq
	s_< =  {4 \over 3}  {\rho_< \over T_<}  =
 {4 \over 3} (2 ) ( {\pi^2 \over 30} ) T^3_<	
\eeq
and by conservation of entropy $s_< =  s_>$  and
\beq
	(T_</T_>)^3  = 11/4 
\eeq
Thus, the photon background is at higher temperature 
than the neutrinos because the $e^{\pm}$
  annihilation energy could not be shared among the neutrinos, and
\beq
	T_\nu = (4/11)^{1/3}   T_\gamma   \simeq 1.9K 
\eeq

As we will see below, standard BBN depends on one single parameter,
the baryon-to-photon ratio, $\eta = n_B/n_\gamma$.  In fact this 
quantity is really related to the net baryon density $n_B - n_{\bar B}$.
It would seem however, that today, $n_{\bar B} = 0$, and that
$ n_B/n_\gamma \sim 2 \times 10^{-10}$. In the absence of baryon number
violation, this ratio is roughly constant (in an adiabatically expanding 
Universe, the ratio of the baryon density to entropy density is
constant).  The question we face therefore, is what determines these
initial conditions, rather than say, $n_B = n_{\bar B}$ and/or
$n_B/n_\gamma = 1$.

Let us for the moment assume that in fact  $\eta  = $  0.  We can 
compute the final number density of nucleons 
left over after annihilations of baryons and 
antibaryons  have frozen out.  At very high 
temperatures, $T  
>$  1 GeV (but neglecting the quark-hadron
 transition), nucleons were in thermal equilibrium with 
the photon background and $n 
_{B} = n_{\bar B} = (3/2)n_\gamma$  (a factor of 2 accounts
 for neutrons and protons and the factor 3/4 
 for the difference between fermi and bose statistics). 
 As the temperature fell below $m_N$,
  annihilations kept the nucleon density at its equilibrium value 
$(n_B/n_\gamma) \simeq (m_{N}/T)^{3/2} {\rm exp}(-m_{N}/T)$ 
 until the annihilation rate  
$\Gamma_A \simeq n_B m_\pi^{-2}$ 
 fell below the expansion rate. This occurred at $T  
\simeq$  20 MeV.  However, at this time the nucleon 
number density had already dropped to
\beq
n_B/n_\gamma = n_{\bar B}/n_\gamma \simeq 10^{-18}
\eeq
 which is eight orders of magnitude too small \cite{Gary} in addition to 
the problem of having to separate the baryons from the antibaryons.
 If any separation did occur at higher temperatures 
(so that annihilations were 
as yet incomplete) the maximum distance scale on which separation could occur 
is the causal scale related to the age of the Universe at that time.  At $T  
=$  20 MeV, the age of the Universe was only $t  =  2 \times 10^{-3}$  
sec.  At that time, a causal region (with distance scale defined by 2ct) 
could only have contained 
$10^{-5} M_\odot$  which is very far from the galactic mass scales on
which we are asking for separations to occur,  
$10^{12} M_\odot$. In spite of all of these problems, $\eta = 0$ implies that
the Universe as a whole is baryon symmetric, thus unless baryons are
separated on extremely large (inflationary) domains, in which case we might
just as well worry again about $\eta \ne 0$, there should be 
antimatter elsewhere in the Universe.  To date, the only antimatter
observed is the result of a high energy collision, either in an 
accelerator or in a cosmic-ray collision in the atmosphere. There has been no
sign to date of any primary antimatter, such as an 
anti-helium nucleus ${\bar \alpha}$
found in cosmic-rays.

The production of a net baryon asymmetry requires baryon number violating
interactions, C and CP violation and a departure 
from thermal equilibrium\cite{sak}.
The first two of these ingredients are contained in
Grand Unified Theories (GUTs), 
the third can be realized in an expanding universe
 where as we have seen, it is not uncommon that interactions 
come in and out of equilibrium.  
In SU(5), the fact that quarks and leptons are in the same multiplets allows
 for baryon non-conserving interactions such as 
$e^{-} + d  \leftrightarrow {\bar u} + {\bar u}$,  etc., 
or decays of the supermassive
 gauge bosons X and Y such as 
$ X  \rightarrow e^{-} + d, {\bar u} + {\bar u}$. 
 Although today these interactions 
are very ineffective because of the very large masses of the X 
and Y bosons, in the early Universe when   
$T \sim M_{ X} \sim 10^{15}$  GeV these types of interactions 
should have been very important.
 C and CP violation is very model dependent.  In the minimal SU(5) 
model
the magnitude of C and CP violation is too small to yield a useful value of  
$\eta$.  The C and CP violation in general  comes 
from the interference between
 tree level decay diagrams and their one loop corrections.

\begin{figure}[htbp]
\hspace{0.5truecm}
\epsfysize=5.5truein\epsfbox{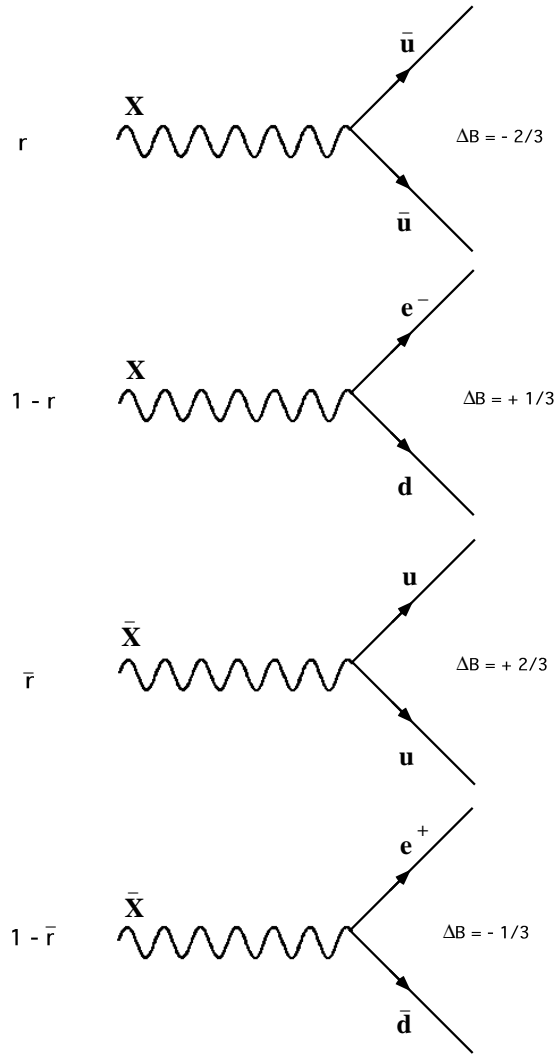}
\caption{{The baryon number violating decays of $X$ and ${\bar X}$.}}
\end{figure}

The departure from equilibrium is very common in the 
early Universe when interaction 
rates cannot keep up with the expansion rate.  In fact, 
the simplest (and most useful) 
scenario for baryon production makes use of the fact that a 
single decay rate goes out of equilibrium.  It is commonly referred to 
 as the out of equilibrium decay scenario 
\cite{ww}.  The basic idea is that the gauge bosons
 $X$ and $Y$ (or Higgs bosons)
 may have a lifetime long enough to insure that the 
inverse decays have already 
ceased so that the baryon number is produced by their free decays.

More specifically, let us call $X$ either the gauge 
boson or Higgs boson, which produces 
the baryon asymmetry through decays.  Let  
$\alpha$  be its coupling to fermions.  For $X$ a gauge boson,  $\alpha$ 
will be the GUT fine structure constant, while for $X$ a Higgs boson,  
$(4{\pi \alpha })^{ 1/2}$ is its Yukawa coupling to fermions. 
 The decay rate for $X$ will be  
\beq
 \Gamma_{ D}  \simeq   \alpha M_{X}
\eeq
  However decays can only begin occurring when the age 
of the Universe is longer
 than the $X$ lifetime   
$\Gamma_D^{-1}$,  i.e., when  $\Gamma_{ D} > $ H  
\beq
  \alpha M_{ X}  \ga  N(T)^{ 1/2} T^2/M_{ P} 
\eeq
 or at a temperature 
\beq
 T^{ 2}  \la  \alpha M_{ X}M_{ P}N(T)^{ -1/2}.
\eeq
Scatterings on the other hand proceed at a rate  
$\Gamma_{ S}  \sim \alpha^2 T^3/M_X^2$ 
 and hence are not effective at lower temperatures.  To be in equilibrium,
decays must have been effective as T fell below 
$M_{ X}$  in order to track the equilibrium 
density of $X$'s (and  ${\bar X}$'s). 
Therefore, the out-of-equilibrium condition  is 
that at $T = M_{ X},   \Gamma {}_{ D} < H$  
or
 \beq
M_{ X} \ga  \alpha M_{ P} (N(M_{ X}))^{ -1/2}  
\sim 10^{18} \alpha {\rm GeV} 
\label{mxmin}
\eeq
 In this case, we would expect a maximal net baryon asymmetry to be produced.

In the out-of-equilibrium decay scenario \cite{ww}, 
the total baryon asymmetry
produced per X, ${\bar X}$ pair is proportional to 
$\Delta B = ({\bar r} - r)$ as can easily be seen from Figure 2.
Here $r$ (${\bar r}$) is the relevant branching ratio for the decay of
 X (${\bar X}$). C and CP violation translate into the condition that
$ r \ne {\bar r}$. 
If decays occur out-of-equilibrium, then at the time of decay,
$n_X \approx n_\gamma$ at $T < M_X$. We then have 
\beq
{n_B \over s} = {(\Delta B) n_X \over s} \sim 
{(\Delta B) n_X \over N(T) n_\gamma} \sim 10^{-2}(\Delta B)
\label{nbmax}
\eeq

Some comments regarding the baryogenesis mechanism described above
are in order:  Initial conditions are not important. Equilibrium
prior to decay erases any net baryon number. The out of equilibrium 
decays of  X and Y then generate a new asymmetry. Therefore, the final 
resulting asymmetry is always the same, given by say, (\ref{nbmax}).
While $n_B/s \sim 10^{-10}$ is possible, the final result is strongly
model dependent.  I should add that several other mechanisms for 
baryogenesis are known to exist \cite{osch}, but all require the same
three key ingredients.

There are several long standing cosmological problems whose solution
we associate with inflation (see refs. 2 and 8). Here I will 
describe only one of these.
The curvature problem (or flatness problem or age problem) 
can manifest itself in
several ways. For a radiation
dominated gas, the entropy density $s \sim T^3$  and 
conservation of entropy implies $R \sim T^{-1}$. 
 Thus assuming an adiabatically expanding
Universe, the quantity ${\hat k} = k/R^2T^2$ is a dimensionless constant.
  If we now
apply the limit in Eq. (\ref{range}) to Eq. (\ref{o-1}) we find
\beq
{\hat	k} = {k \over R^2 T^2}  = {(\Omega_o  - 1){H_o}^2 \over {T_o}^2}
  < 2 \times 10^{-58}
\eeq
This limit on $k$ represents an initial condition on the cosmological model.
The problem is: what physical processes in the early Universe
produced a value of $\hat{k}$ so extraordinarily close to zero 
(or $\Omega$ close to one)?
	A more natural initial condition might have 
been $\hat{k} \sim 0(1)$.  In this
case the Universe would have become curvature 
dominated at $T \sim 10^{-1} M_P$.

It is important to note that $\Omega$ is a function of 
time or of the scale factor.
The evolution of $\Omega$ is shown in Figure 3 for $\Lambda = 0$.
\begin{figure}
\hspace{0.5truecm}
\epsfysize=7truecm\epsfbox{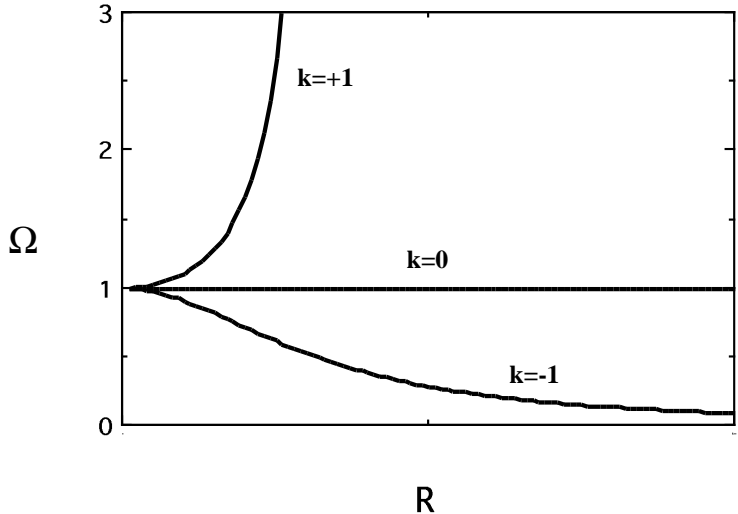}
\caption{The evolution of the cosmological density parameter, $\Omega$,
as a function of the scale factor for a closed, open and 
spatially flat Universe.}
\label{fig:omegaf}
\end{figure}
For a spatially flat Universe, $\Omega = 1$ always. When $k = +1$,
there is a maximum value for the scale factor $R$. At early
times (small values of $R$), $\Omega$ always tends to one.
Note that the fact that we do not yet know the sign of
$k$, or equivalently whether $\Omega$ is larger than or smaller than
unity, implies that we are at present still at the very left in the
figure.  What makes this peculiar is that one would normally expect that
the sign of $k$ to become apparent after a Planck time of $10^{-43}$ s.
It is extremely puzzling that some $10^{60}$ Planck times later,
we still do not know the sign of $k$.

The curvature problem discussed above  among others 
can be neatly resolved if the 
Universe underwent a period of cosmological inflation \cite{infl}.
 During a phase transition, our assumptions of an
adiabatically expanding universe may not be valid.  
If we look at a scalar potential 
describing a phase transition from a symmetric false vacuum state
 $\langle \phi \rangle  = 0$ for some 
scalar field $\phi$ to the broken true vacuum at
 $\langle \phi \rangle  = v$, 
and suppose we find that upon solving the equations of motion for the
scalar field that the field evolves slowly from the symmetric
state to the global minimum (this will depend on the details of the 
potential).  If the evolution is slow enough, the universe may become dominated
by the vacuum energy density associated with 
the potential near $\phi \approx 0$.
The energy density of the symmetric vacuum, V(0) acts as a cosmological 
constant with
\beq
	\Lambda = 8\pi V(0) {M_P}^2	
\eeq
During this period of slow evolution, the energy 
density due, to say, radiation will fall below the vacuum 
energy density, $\rho \ll V(0)$.  When this happens, the expansion 
rate will be dominated by the constant V(0) and from Eq. (\ref{H})
we find an exponentially expanding solution
\beq
R(t) \sim e^{\sqrt{\Lambda/3}~t}
\label{DS}
\eeq
When the field evolves towards the global minimum 
it will begin to oscillate
about the minimum, energy will be released
during its decay and a hot thermal universe will be restored.
 If released 
fast enough, it will produce radiation at a temperature 
${T_R}^4 \la V(0)$.  In this reheating process entropy has been created and
	$(RT)_f  > (RT)_i $.	
Thus we see that during a phase
 transition the relation $RT \sim$ constant, need not hold true and thus 
our dimensionless constant $\hat{k}$ may actually not have been constant.

If during the phase transition,  the value of
 $RT$ changed by a factor of $0(10^{29})$, the cosmological
 problems would be solved.  The isotropy would in a sense be
 generated by the immense expansion; one small causal region
 could get blown up and hence our entire visible Universe would 
have been at one time in thermal contact.  In addition, 
the parameter $\hat{k}$ could have started out $0(1)$ and have
 been driven small by the expansion.

\section{Big Bang Nucleosynthesis}

The concordance between big bang nucleosynthesis (BBN) theory and observation
has been the subject of considerable recent debate.  It is clear however,
that the real questions lie not with the concordance between BBN and 
the observational data, but rather between the theories of chemical
and stellar evolution and the data.  BBN theory (see for example ref. 9)
is quite stable in the sense that over time very little
in the fundamental theory
has changed.  Cross-sections are now more accurately
measured, the neutron mean life is known with a much higher degree of
precision, and if we restrict our attention to the standard model, the number
of neutrinos has also been determined. In contrast, the status of the 
observational data has changed significantly in the last several years.
There is better data on \he4, more data on \li7, and data on D and \he3 
that was simply non-existent several years ago.  For the most part, the
inferred abundances of \he4 and \li7 have remained relatively fixed, giving us 
a higher degree of confidence in the assumed primordial abundances of these
isotopes as is reflected in their observational uncertainties. Indeed, the
abundances of \he4 and \li7 alone are sufficient in order to probe and 
test the theory and determine the single remaining parameter in the standard
model \cite{fo}, namely, the baryon-to-photon ratio, $\eta$.
In contrast, D and \he3 are highly dependent on models of chemical evolution
(\he3 is in addition dependent on the uncertain stellar yields of this isotope).
New data from quasar absorption systems, on what may be primordial D/H is at
this time disconcordant, different measurements give different abundances.
As a consequence of the uncertainties in D and \he3, 
one can use the predictions based on \he4 and \li7 in order
to construct models of galactic chemical evolution.  These results also have 
important implications for the amount of (non)-baryonic dark matter in the
galaxy and on the number of allowed relativistic degrees of freedom at the
time of BBN, commonly parameterized as $N_\nu$. The basic results of BBN
are summarized in Figure 4, showing the range in $\eta$ where consistency 
with the observations is achieved.

\begin{figure}[htbp]
\hspace{0.5truecm}
\epsfysize=5.5truein\epsfbox{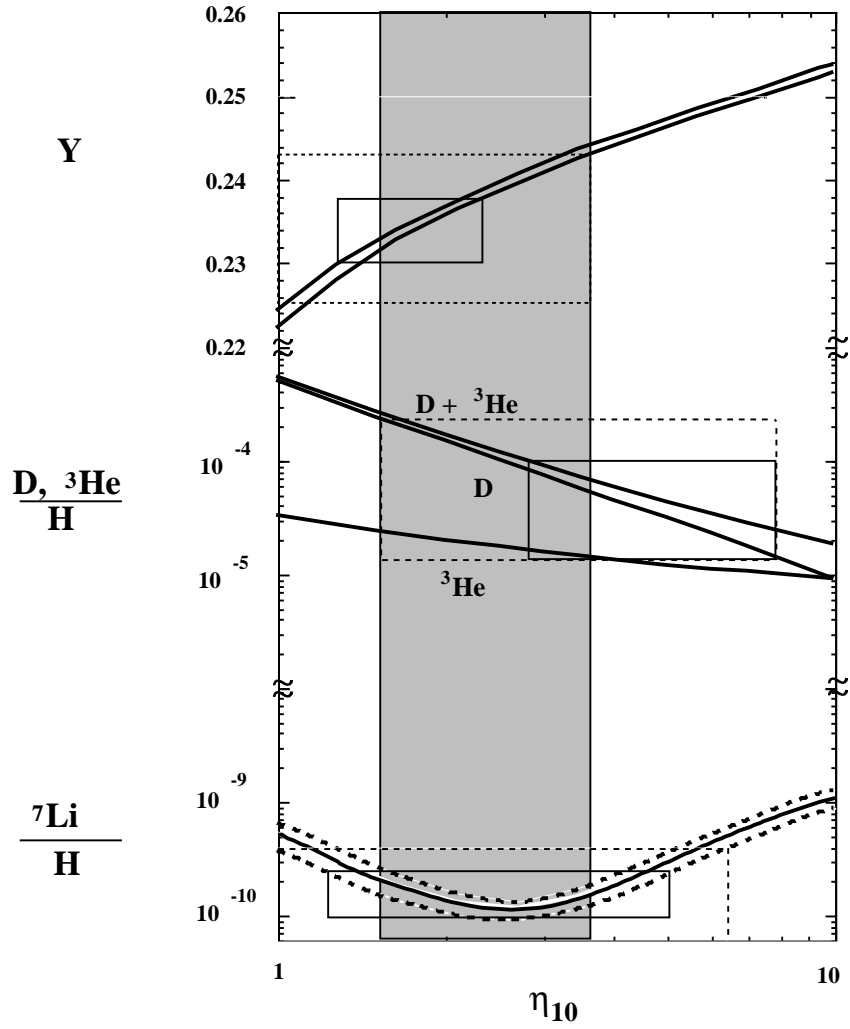}
\caption{{The light element abundances from big bang
nucleosynthesis as a function of $\eta_{10} = 10^{10}\eta$.}}
\end{figure}

There has always been an intimate connection between BBN and the microwave
background as key test to the standard big bang model.
Indeed, it was the formulation of BBN which predicted the existence of 
the microwave background radiation \cite{gamo}. 
The argument is rather simple. BBN requires temperatures greater than
100 keV, which according to eqs. (\ref{tt1}) and (\ref{tt2}) corresponds
to timescales less than about 200 s. The typical cross section for the
first link in the nucleosynthetic chain is
\beq 
\sigma v (p + n \rightarrow D + \gamma) \simeq 5 \times 10^{-30} 
{\rm cm}^3/{\rm s}
\eeq
This implies that it was necessary to achieve a density
\beq
n \sim {1 \over \sigma v t} \sim 10^{17} {\rm cm}^{-3}
\eeq
Now the density in baryons today is known approximately from the density of
visible matter to be ${n_B}_o \sim 10^{-7}$ cm$^{-3}$ and since
we know that that the density $n$ scales as $R^{-3} \sim T^3$, 
the temperature today must be
\beq
T_o = ({n_B}_o/n)^{1/3} T_{\rm BBN} \sim 10 {\rm K}
\eeq
A pretty good estimate.  

Despite its simplicity, BBN was criticized early on, due to its 
shortcomings in being able to produce the observed abundances of 
{\em all} of the element isotopes.  Attention was therefore turned to
stellar nucleosynthesis.  However, while the elements from helium on up
can be and are produced in stars, no other astrophysical site has ever
survived for the production of deuterium.  In addition, if one assumes that
\he4 is entirely of stellar origin, one should be able to find places in the 
Universe in which the \he4 mass fraction is substantially below 25\%.
The  \he4 data shown in Figure 5, emphasizes the fact that indeed 
no such region with low \he4 has ever been observed and that (together with
the need to produce D) leads one to conclude that BBN nucleosynthesis 
is a necessary component in any cosmological model.

\begin{figure}
\hspace{0.5truecm}
\epsfysize=7truecm\epsfbox{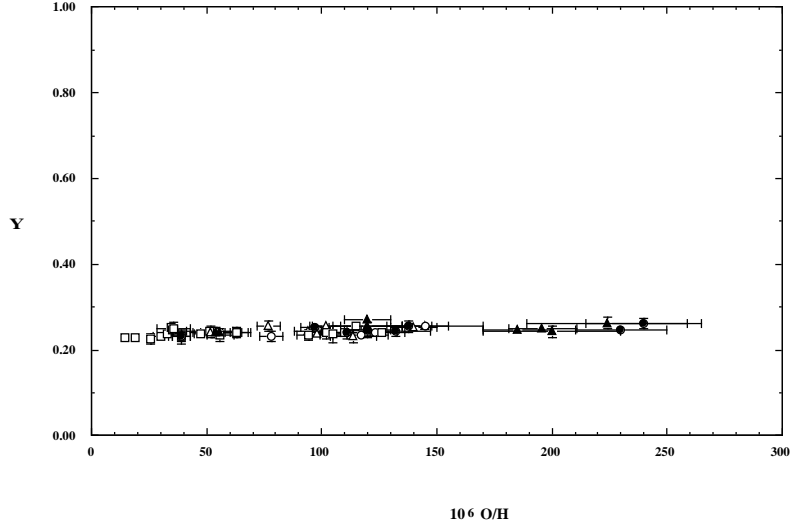}
\caption { The helium (Y) vs oxygen (O/H) abundances 
in extragalactic HII regions emphasizing the lack of low \he4 regions.  }
\label{fig:fig4gary}
\end{figure}

Before commencing with the direct comparison between theory and observations,
it will be useful to briefly review the main events leading to the synthesis
of the light elements.
Conditions for the synthesis of the light elements were attained in the
early Universe at temperatures  $T \la $ 1 MeV.  
At somewhat higher temperatures, weak interaction rates were
in equilibrium, thus fixing the ratio of number densities of neutrons to
protons. At $T \gg 1$ MeV, $(n/p) \simeq 1$.  As the temperature fell and
approached the point where the weak interaction rates were no longer fast enough
to maintain equilibrium, the neutron to proton ratio was given approximately by
the Boltzmann factor, $(n/p) \simeq e^{-\Delta m/T}$, where $\Delta m$
is the neutron-proton mass difference. The final abundance of \he4 is very
sensitive to the $(n/p)$ ratio.

The nucleosynthesis chain begins with the formation of deuterium
through the process, $p+n \rightarrow$ D $+ \gamma$.
However, because the large number of photons relative to nucleons,
$\eta^{-1} = n_\gamma/n_B \sim 10^{10}$, deuterium production is delayed past
the point where the temperature has fallen below the deuterium binding energy,
$E_B = 2.2$ MeV (the average photon energy in a blackbody is
${\bar E}_\gamma \simeq 2.7 T$).  
When the quantity $\eta^{-1} {\rm exp}(-E_B/T) \sim 1$ the rate for 
deuterium destruction (D $+ \gamma \rightarrow p + n$)
finally falls below the deuterium production rate and
the nuclear chain begins at a temperature $T \sim 0.1 MeV$.
The nuclear chain in BBN  calculations was extended \cite{tsof} and
is shown in Figure 6.

\begin{figure}[htbp]
\hspace{0.5truecm}
\epsfysize=6.0truein\epsfbox{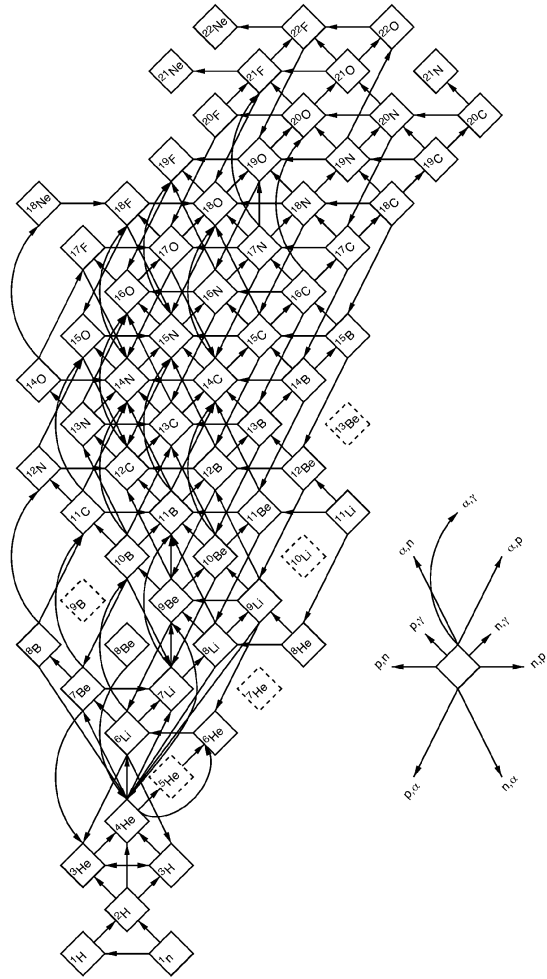}
\caption{{The nuclear network used in BBN calculations.}}
\end{figure}

The dominant product of big bang nucleosynthesis is \he4 resulting in an
abundance of close to 25\% by mass. This quantity is easily estimated by
counting the number of neutrons present when nucleosynthesis begins.
When the weak interaction rates responsible for $n-p$
equilibrium freeze-out, at $T \approx 0.8$ MeV,
the neutron to proton ratio is about 1/6. When free neutron decays 
 prior to deuterium formation are taken into account, the ratio drops to
$(n/p) \approx 1/7$. Then simple counting yields a \he4  mass fraction
\beq
Y_p = {2(n/p) \over \left[ 1 + (n/p) \right]} \approx 0.25
\label{ynp}
\eeq

In the standard model,
 the \he4 mass fraction
depends primarily on the baryon to photon ratio,
$\eta$ as it is this quantity which determines the onset of nucleosynthesis 
via deuterium production. 
For larger values of $\eta$, the quantity 
$\eta^{-1} {\rm exp}(-E_B/T) \sim 1$ is smaller, and hence the nuclear chain
may begin at a higher temperature.  As a result, the $(n/p)$ ratio is 
higher, producing more \he4.
But because the $(n/p)$ ratio is only
weakly dependent on $\eta$, the \he4 mass fraction is relatively
flat as a function of $\eta$. When we go beyond the standard model, the
\he4 abundance is very sensitive to changes in the expansion rate which 
can be related to the effective number of neutrino flavors as will
be discussed below. Lesser amounts of the other light elements are produced:
D and \he3 at the level of about $10^{-5}$ by number, 
and \li7 at the level of $10^{-10}$ by number. These abundances (along with
\li6) are shown in Figure 7 \cite{tsof}.  In Figure 8, 
the produced abundances of the intermediate mass isotopes \be9, \b10,
\b11 are also shown.  These abundances are far below the observed values
and it is believed that these isotopes are formed in cosmic
ray nucleosynthesis.

\begin{figure}[htbp]
\hspace{0.5truecm}
\epsfysize=5.0truein\epsfbox{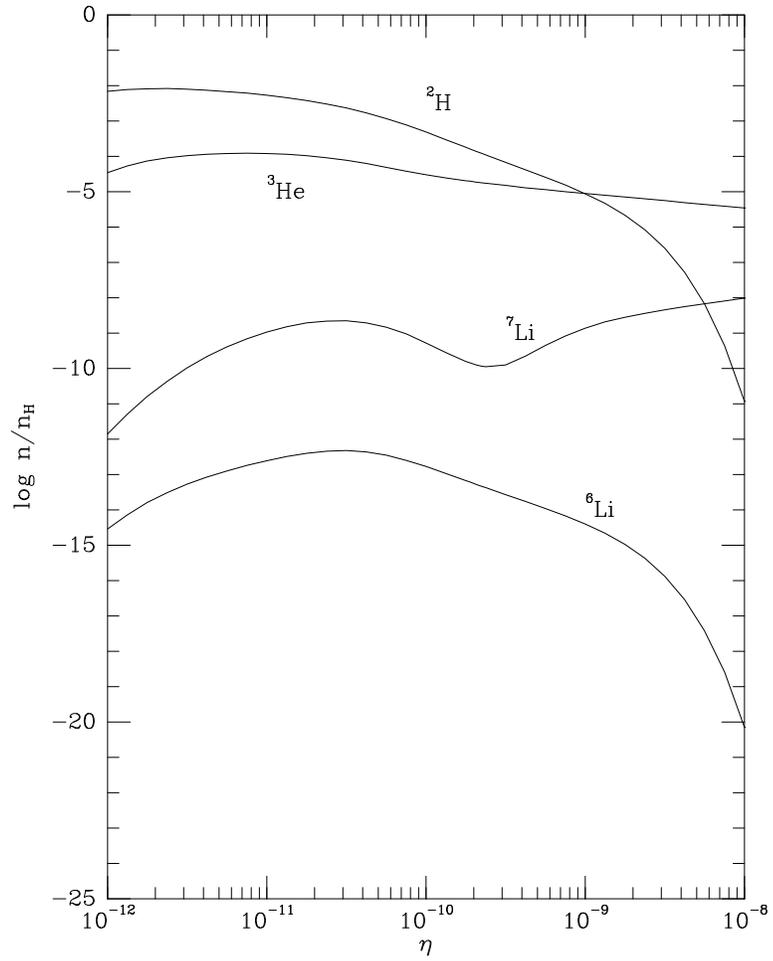}
\caption{{The light element abundances from big bang
nucleosynthesis as a function of $\eta$, including \li6.}}
\end{figure}
\begin{figure}[htbp]
\hspace{0.5truecm}
\epsfysize=5.0truein\epsfbox{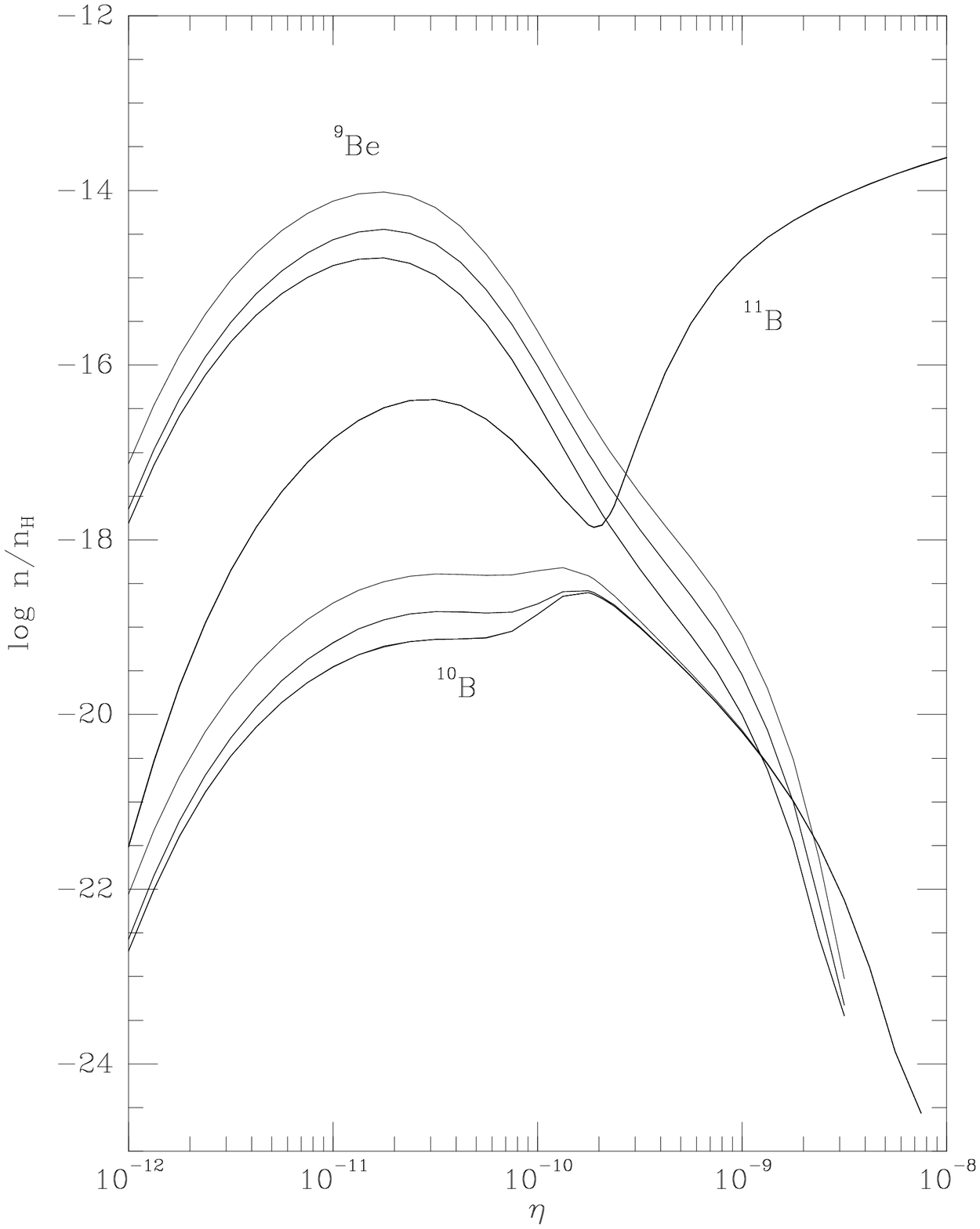}
\caption{{The intermediate mass element abundances from big bang
nucleosynthesis as a function of $\eta$.}}
\end{figure}

For the comparison with the observations, I will use
the resulting abundances of the light elements shown in Figure 4. 
The curves for the \he4 mass fraction, $Y$, bracket the
computed range based on the uncertainty of the neutron mean-life which 
has been taken as \cite{rpp} $\tau_n = 887 \pm 2$ s. 
 Uncertainties in the produced \li7 
abundances have been adopted from the results in Hata et al.
\cite{hata1}. Uncertainties in D and
\he3 production are small on the scale of this figure. 
The  boxes correspond
to the observed abundances and will be discussed below. 

\subsection{Light Element Abundances}

There is now a good collection of abundance information on the \he4 mass
fraction, $Y$, O/H, and N/H in over 70 extragalactic HII 
(ionized hydrogen) regions
\cite{p,evan,iz}. 
The \he4 abundance in very low metallicity regions is best determined
from observations  of HeII $\rightarrow$ HeI
recombination lines. The observation of the heavy elements is 
important as the helium
mass fraction observed in these HII regions has been 
augmented by some stellar
processing, the degree to which is 
given by the oxygen and nitrogen abundances.
In an extensive study based on the data in \cite{p,evan}, it was found
\cite{osa}   that the data is well represented by a linear correlation for
$Y$ vs. O/H and Y vs. N/H. It is then expected that the primordial abundance
of \he4 can be determined from the intercept of that relation.  
The overall result
of that analysis indicated a primordial mass fraction, 
 $Y_p  = 0.232 \pm 0.003$.
The stability of this fit was verified by a 
statistical bootstrap analysis \cite{osc}
showing that the fits were not overly sensitive to any particular HII region.

To make use of the \he4 data, it is crucial to obtain
high quality and very low metallicity data.
In principle,  any one HII region (with 
non-zero metallicity) should provide an upper limit to $Y_p$
since some stellar processing has taken place augmenting the primordial
value. Thus the determination of $Y_p$ by an extrapolation to zero 
metallicity could be avoided by the observations of either 
low metallicity or low helium HII regions.  
For a very low metallicity HII region such an upper limit may even 
provide a reasonable estimate of $Y_p$. 
For example, I Zw 18, is the lowest metallicity extragalactic 
HII region with \he4 observed.  In I Zw 18, there are 2 separate regions,
and there have been five distinct measurements of \he4.  In the HII 
regions designated 
as NW and SE, the five measured abundances are shown in the table below.
\begin{table}[ht]
\centerline {\sc{\underline{Table 1:} Data for I Zw 18}}
\begin{center}
\begin{tabular}{|ccc|}                    \hline
NW & $.216 \pm .012$ & ref. 15 \\ 
NW & $.231 \pm .006$ & ref. 16 \\ 
NW & $.232 \pm .008$ & ref. 17 \\ 
SE & $.231 \pm .012$ & ref. 15 \\ 
SE & $.230 \pm .008$ & ref. 16 \\ 
\hline
\end{tabular}
\end{center}
\end{table}
This data leads to a total average for I Zw 18 of $Y_p = 0.230 \pm 0.004$.
There are several other regions with low \he4 abundances and  
the value of Y 
derived from the average of several HII regions is also of interest \cite{ost3}.  
If we successively include additional HII regions with 
higher Y, as more regions are included, the mean value 
(weighted) of Y will increase, but if the errors are statistical, the 
error in the mean will decrease.  As a result, for $N$ HII regions the 
one-(or two-)$\sigma$ upper bound to Y will first decrease with $N$, 
then level off and, as $N$ is further increased, it will eventually 
increase monotonically.  This behavior is seen in Figure 9 where  
weighted means are shown, and the 1$\sigma$ bounds to the 
weighted means of Y derived from the $N$ lowest helium abundance HII 
regions.  
Note that for $2 \leq N \leq 13$, the mean varies from 0.229 to 
0.231 
while for $2 \leq N \leq 14$, $\langle Y \rangle \leq 
0.236$(2$\sigma$).  
It is not unreasonable to infer from these results that,
\beq
Y_p \leq 0.230\pm 0.003
\eeq
with, $Y_{p}^{2\sigma} \leq 0.236$.  If, instead, we take the weighted
means of the regions with the lowest values of O/H, we 
obtain a similar result.  This illustrates the potentially great 
value of very careful analyses of a handful of the lowest metallicity 
(lowest Y) HII regions. 

\begin{figure}
\hspace{0.5truecm}
\epsfysize=7truecm\epsfbox{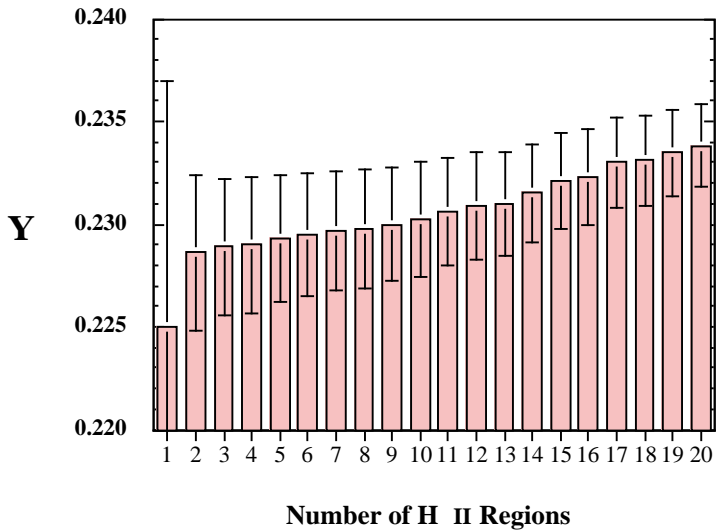}

\caption { \baselineskip=2ex The running average (weighted means) of the 
\he4 abundance, $Y$, 
for the first $N$ (lowest $Y$) HII regions (from ref. 20).  Also shown are the 
1$\sigma$ bounds to the weighted means. }
\label{fig:figost35}
\end{figure}

Although the above estimates on $Y_p$ are consistent with those based on 
a linear extrapolation of the data in refs. 15 and 16,
it has been claimed, that the new data in ref. 17,
leads to a significantly higher value for $Y_p$ (in excess of 24\%).
However, in a recent analysis \cite{ost3} it has been shown that
the data in ref. 17 is entirely consistent with the data in
refs. 15 and 16 as can be seen in Figure 10 where 
the Y versus O/H data used in ref. 18 (open circles, based on the data in
refs. 15 and 16)) is shown along with the new 
data (filled circles, from ref. 17).  
The crossed circles are the 10 HII regions 
that Izotov et al.  excluded from their  own analyses.  
Note that where there is overlap 
in O/H, the newer Y values are intermingled with those from refs 15 and 16.  
It was argued \cite{ost3} that because the data in ref. 17 does not extend down
to sufficiently low metallicity, taken alone it can not be used to determine 
$Y_p$.  It was also shown \cite{ost3} that one can not argue for a systematic
shift in $Y$ on the basis of new atomic data calculations as was done in
ref. 17.
On the other hand, this data may be combined with the 
previous data in which case one finds
\begin{figure}
\hspace{0.5truecm}
\epsfysize=7truecm\epsfbox{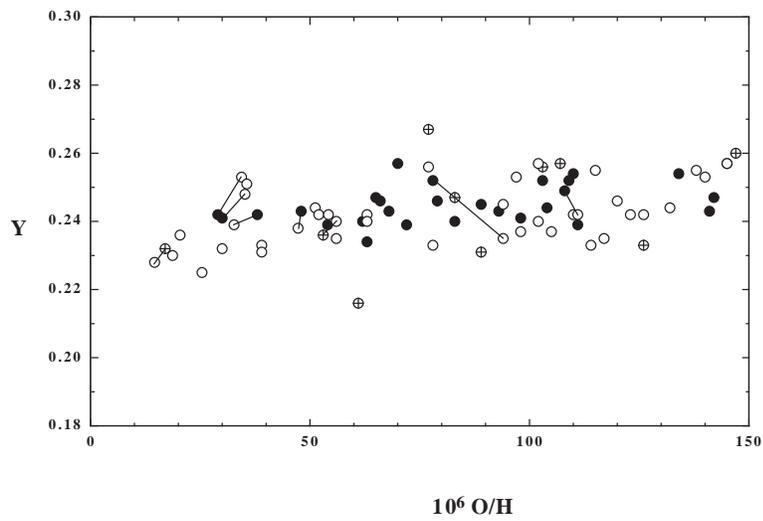}

\caption { \baselineskip=2ex The helium (Y) and oxygen (O/H) abundances 
from ref. 20 in extragalactic HII regions, 
(open circles) from refs. 15 and 16, and from ref. 17 (filled circles).  
Regions excluded by ref. 17 are shown as crossed circles.  Lines connect the 
same regions observed by different groups.  }
\label{fig:figost36}
\end{figure}
a \he4 mass fraction \cite{ost3} based on 62 distinct HII regions
\beq
Y_p = 0.234 \pm 0.002 \pm 0.005
\label{he4}
\eeq
The first uncertainty is purely statistical and
the second uncertainty is an estimate of the systematic uncertainty in the
primordial abundance determination. The solid box for \he4 in Figure 4
represents the range (at 2$\sigma_{\rm stat}$)
from (\ref{he4}). The dashed box extends this by $\sigma_{\rm sys}$.
A
somewhat lower primordial abundance of  $Y_p = 0.230 \pm .003$ is found
by restricting to the 32 most metal poor regions \cite{ost3}.  


The \li7 abundance
is also reasonably well known.
 In old,
hot, population-II stars, \li7 is found to have a very
nearly  uniform abundance. For
stars with a surface temperature $T > 5500$~K
and a metallicity less than about
1/20th solar (so that effects such as stellar convection may not be important),
the  abundances show little or no dispersion beyond that which is
consistent with the errors of individual measurements.
Indeed, as detailed in ref. 21, much of the work concerning
\li7 has to do with the presence or absence of dispersion and whether
or not there is in fact some tiny slope to a [Li] = $\log$ \li7/H + 12 vs.
T or [Li] vs. [Fe/H]  relationship ([Fe/H] is the log of the Fe/H ratio
relative to the solar value).

There is \li7 data from nearly 100 halo stars, from a 
 variety of sources. When the Li data from stars with [Fe/H] $<$ 
-1.3 is plotted as a function of surface temperature, one sees a 
plateau emerging for $T > 5500$ K as shown in Figure 11 for 
the data taken from ref. 22.
\begin{figure}
\hspace{0.5truecm}
\epsfysize=7truecm\epsfbox{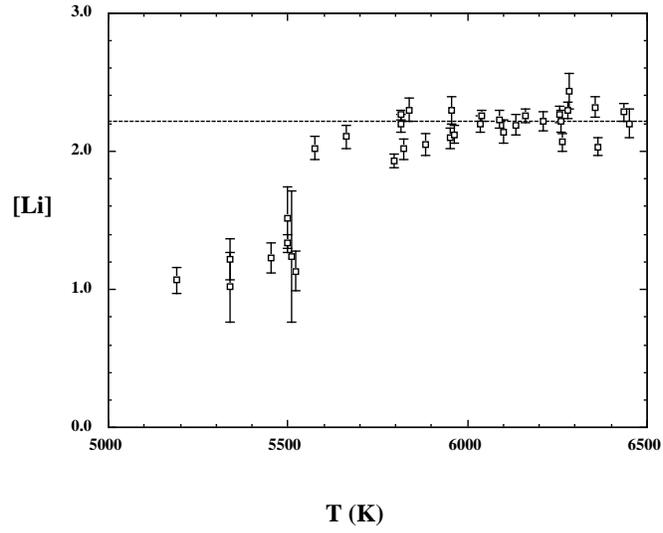}
\caption{The Li abundance in halo stars with [Fe/H] $<$ -1.3,
as a function of surface temperature}
\end{figure}
As one can see from the figure, at high temperatures, where the convection
zone does not go deep below the surface, the Li abundance is uniform.
At lower temperatures, the surface abundance of Li is depleted as
Li passes through the interior of the star and is destroyed.
The lack of dispersion in the plateau region is evidence that this abundance
is indeed primordial (or at least very close to it).  
Another way to see the plateau is to plot the Li abundance data as a 
function of metallicity, this time with the restriction that
$T > 5500$ K as seen in Figure 12.
\begin{figure}
\hspace{0.5truecm}
\epsfysize=7truecm\epsfbox{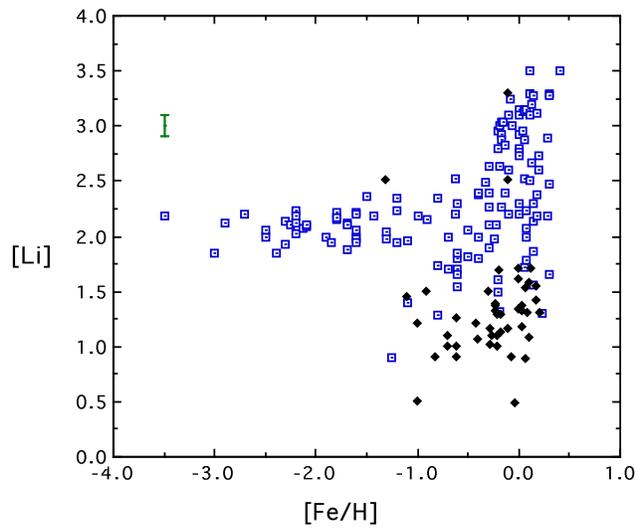}
\caption{The Li abundance in halo stars with $T > 5500$ K,
as a function of metallicity. Filled diamonds represent upper limits.}
\end{figure}
In this case, the plateau emerges at low [Fe/H] as would be expected.
At higher [Fe/H], the convection zone remains below the surface 
only for much hotter stars. Thus, for [Fe/H] $>$ -1.3, the effects
of depletion are seen.  Also apparent in this figure is that at higher
metallicities there is evidence for the production of Li which
rises by over an order of magnitude at solar metallicity.

I will use the value given in ref. 22 
as the best estimate
for the mean \li7 abundance and its statistical uncertainty in halo stars 
\beq
{\rm Li/H = (1.6 \pm 0.1 {}^{+0.4}_{-0.3} {}^{+1.6}_{-0.5}) \times 10^{-10}}
\label{li}
\eeq
 The first error is statistical, and the second
is a systematic uncertainty that covers the range of abundances
derived by various methods. 
The Li abundance is somewhat sensitive to stellar parameters such as
the assumed surface temperature, the metallicity and the surface gravity.
The greatest model dependence is on the conversion of the observed
colors (B-V) to temperature.  For example, in a sample of 55 stars taken
from the papers of ref. 23, one finds [Li] = $2.08 \pm 0.02$.
From Thorburn's\cite{thor} sample of 74 stars one finds [Li] = $2.26 \pm
0.01$.  And ref. 22, which comprises 41 stars gives [Li] = $2.21 
\pm 0.01$. The Li abundance used in (\ref{li}) is derived from this
last data set.  
The solid box for \li7 in Figure 4 represents
the 2$\sigma_{\rm stat} + \sigma_{\rm sys}$ range from (\ref{li}).

The third set of errors deals with the possible depletion and/or
production of \li7.
These uncertainties are however limited.  As was noted above, the lack
of dispersion in the Li data limits the amount of depletion. 
In addition,  standard stellar
models\cite{del} predict that any depletion of \li7 would be 
accompanied by a
very severe depletion of \li6.  Until recently, 
\li6 had never been observed in
hot pop II stars. The observation\cite{li6o} of 
\li6 (which turns out to be
consistent with its origin in cosmic-ray nucleosynthesis 
and with a small amount
of depletion as expected from standard stellar models) 
is  another good indication that
\li7 has not been destroyed in these stars\cite{li6}.

Aside from the big bang, Li is produced together with Be and B in cosmic ray
spallation of C,N,O by protons and $\alpha$-particles.  Li is also produced by 
 $\alpha-\alpha$ fusion.  Be and B have been observed in these same
pop II stars and in particular there are a dozen or so stars in which both
Be and \li7 have been
observed.  Thus Be (and B though there is still a paucity of
data) can be used as a consistency check on primordial Li \cite{fossw}. Based on
the Be abundance found in these stars, 
one can conclude that no more than 10-20\% of 
the \li7 is due to cosmic ray nucleosynthesis leaving the remainder
(an abundance near $10^{-10}$) as primordial.
The third set of errors in Eq. (\ref{li}) accounts for
 the possibility that as much as half
of the primordial \li7 has been
destroyed in stars, and that as much as 20\% of the observed \li7 may have been
produced in cosmic ray collisions rather than in the Big Bang.
The dashed box in Figure 4, accounts for this additional uncertainty.
 For \li7, the uncertainties are clearly dominated by
systematic effects.

Turning to D/H, we have three basic types of abundance information:
1) ISM data, 2) solar system information, and perhaps 3) a primordial
abundance from quasar absorption systems.  The best measurement for ISM D/H
is \cite{linetal}
\beq
{\rm (D/H)_{ISM}} = 1.60\pm0.09{}^{+0.05}_{-0.10} \times 10^{-5}
\eeq
Because there are no known astrophysical sites for the production of
deuterium, all observed D must be primordial. As a result,
a firm lower bound from deuterium establishes an upper bound on $\eta$
which is robust and is shown by the lower right of the solid box
in Figure 4.
 The solar abundance of D/H is inferred from two
distinct measurements of \he3. The solar wind measurements of \he3 as well as 
the low temperature components of step-wise heating measurements of \he3 in
meteorites yield the presolar (D + \he3)/H ratio, as D was 
efficiently burned to
\he3 in the Sun's pre-main-sequence phase.  These measurements 
indicate that \cite{scostv,geiss}
\beq
{\rm \left({D +~^3He \over H} \right)_\odot = (4.1 \pm 0.6 \pm 1.4) \times
10^{-5}}
\eeq
 The high temperature components in meteorites are believed to yield the true
solar \he3/H ratio of \cite{scostv,geiss}
\beq
{\rm \left({~^3He \over H} \right)_\odot = (1.5 \pm 0.2 \pm 0.3) \times
10^{-5}}
\label{he3}
\eeq
The difference between these two abundances reveals the presolar D/H ratio,
giving,
\beq
{\rm (D/H)_{\odot}} \approx (2.6 \pm 0.6 \pm 1.4) \times 10^{-5}
\eeq

It should be noted that recent measurements of surface abundances 
of HD on Jupiter
show a somewhat higher value for D/H,  D/H = $5 \pm 2 \times 10^{-5}$ 
\cite{nie}. If this value is confirmed and if fractionation
does not significantly alter the D/H ratio (as it was suspected to for 
previous measurements involving CH$_3$D), it may have an important 
impact on galactic chemical evolution models.  This value
is marginally consistent with the inferred meteoritic values. 

Finally, there have been several recent reported measurements of 
D/H is high redshift quasar absorption systems. Such measurements are in
principle capable of determining the primordial value for D/H and hence $\eta$,
because of the strong and monotonic dependence of D/H on $\eta$.
However, at present, detections of D/H  using quasar absorption systems
do not yield a conclusive value for D/H.  As such, it should be cautioned 
that these values may not
turn  out to represent the true primordial value and it is very unlikely 
that both are primordial and indicate an inhomogeneity \cite{cos2}.
The first of these measurements \cite{quas1} indicated a rather high D/H ratio,
D/H $\approx$ 1.9 -- 2.5 $\times 10^{-4}$.  Other 
high D/H ratios were reported in \cite{quas3}. 
However, there are reported low
values of D/H in other such systems \cite{quas2} with values D/H $\simeq 2.5
\times 10^{-5}$, significantly lower than the ones quoted above. 
Though this primordial D/H value is consistent with the solar 
and present values
of D/H, it is not consistent (at the 2 $\sigma$ level) with 
the determinations
of D/H in Jupiter, if they are correct. It is generally regarded
that the high D/H measurements are of lesser quality while  
 a recent observation  \cite{swc} of 
the hydrogen column density in one of low D/H clouds found a significantly
 lower value and claimed a lower limit on D/H of $4 \times 10^{-5}$.

The range of quasar absorber D/H is shown by the dashed box in Figure 4.
It is probably
premature to use either of these values as the primordial D/H abundance in 
an analysis of big
bang nucleosynthesis, but it is certainly encouraging that 
future observations may
soon yield a firm value for D/H. It is however important to 
note that there does
seem to be a  trend that over the history of the Galaxy, the D/H ratio  is
decreasing, something we expect from galactic chemical evolution.  
Of course the
total amount of deuterium astration that has occurred is still uncertain, and
model dependent.

There are also several types of \he3 measurements. As noted above, meteoritic
extractions yield a presolar value for \he3/H as given in Eq. (\ref{he3}).
In addition, there are several ISM measurements of \he3 in galactic HII
regions \cite{bbbrw} which show a wide dispersion which may be indicative 
of pollution or a bias \cite{orstv}
\beq
 {\rm \left({~^3He \over H} \right)_{HII}} \simeq 1 - 5 \times 10^{-5}
\eeq
There is also a recent ISM measurement of \he3 \cite{gg}
with
\beq
 {\rm \left({~^3He \over H} \right)_{ISM}} = 2.1^{+.9}_{-.8} \times 10^{-5}
\eeq
  Finally there are observations of \he3 in planetary
nebulae \cite{rood} which show a very high \he3 abundance of 
\he3/H $\sim 10^{-3}$.

Each of the light element isotopes can be made consistent with theory for a
specific range in $\eta$. Overall consistency of course requires that
the range in $\eta$ agree among all four light elements.
However, as will be argued below D and \he3 are far more sensitive to 
chemical evolution than \he4 or \li7 and as such the direct comparison
between the theoretical predictions of the primordial abundances of
D and \he3 with the observational determination of their abundances is far more 
difficult.  Therefore in what follows I will for the most part 
restrict the comparison between
theory and observation to the two isotopes who suffer the least from the
effects of chemical evolution.

\subsection{Chemical Evolution}
Because we can not directly measure the primordial abundances of any of the
light element isotopes, we are required to make some assumptions concerning
the evolution of these isotopes. As has been discussed above, 
\he4 is produced in stars along with oxygen and nitrogen.
\li7 can be destroyed in stars and produced in several
(though still uncertain) environments. D is totally destroyed in the star 
formation process and \he3 is both produced and destroyed in stars with
fairly uncertain yields. It is therefore preferable, if possible
to observe the light element isotopes in a low metallicity 
environment. Such is the case with \he4 and \li7, and we can be fairly
assured that the abundance determinations of these isotopes are close to 
primordial.  If the quasar absorption system measurements of D/H stabilize,
then this too may be very close to a primordial measurement.  Otherwise,
to match the solar and present abundances of D and \he3 to their 
primordial values requires a model of galactic chemical evolution.

 The main inputs to chemical evolution models are:
 1) The initial mass function, $\phi(m)$, indicating the 
distribution of stellar masses. Typically, a simple
power law form for the IMF is chosen, $\phi(m) \sim m^{-x}$,
with $x \simeq -2.7$.  This is a fairly good representation of the
observed distribution, particularly at larger masses.
 2) the star formation rate, $\psi$. Typical choices for a SFR
are $\psi(t) \propto \sigma$ or $\sigma^2$ or even a straight exponential
$e^{-t/\tau}$.  $\sigma$ is just the fraction of mass in gas, 
$M_{\rm gas}/M_{\rm tot}$. 3) the presence
of infalling or outflowing gas; and of course 4) the stellar yields.  It is 
the latter, particularly in the case of \he3, that is the cause for
so much uncertainty. Chemical evolution models simply set up a series of 
evolution equations which trace desired quantities.  For example,
$\sigma$ and hence the SFR evolve through a relation such as 
\beq
{dM_{\rm gas} \over dt} = -\psi(t) + e(t) + i(t) - o(t)
\eeq
where $e$ represents the amount of gas ejected from stars, $i$
is the gas infall rate, and $o$ is the gas outflow rate.
Similar equations can be developed which trace the abundances of the
element isotopes \cite{bt}.

Deuterium is always a monotonically decreasing function of time in chemical
evolution models.  The degree to which 
D is destroyed, is however a model dependent
question which depends sensitively on the IMF and SFR.
The evolution of \he3 is however considerably more complicated.
Stellar models predict that substantial amounts of \he3 are
produced in stars between 1 and 3 M$_\odot$. For M $<$ 8M$_\odot$, Iben and
Truran \cite{it} calculate
\beq
(^3{\rm He/H})_f = 1.8 \times 10^{-4}\left({M_\odot \over M}\right)^2 
+ 0.7\left[({\rm D+~^3He)/H}\right]_i
\label{it}
\eeq
so that for example, when $\eta_{10} = 3$, 
((D + \he3)/H)$_i = 9 \times 10^{-5}$, and the ratio of the final abundance
of \he3/H to the initial (D + \he3)/H abundance denoted by $g_3$ is 
$g_3(1 $M$_\odot$) = 2.7. The \he3 abundance is nearly tripled.
It should be emphasized that this prediction is in
fact consistent with the observation of high \he3/H in planetary nebulae
\cite{rood}.

Generally, implementation of the \he3 yield in Eq. (\ref{it}) in chemical
evolution models leads to an overproduction of \he3/H particularly at the
solar epoch \cite{orstv,galli}.  This problem is compounded in models with
an intense period of D destruction.
In Scully et al. \cite{scov}, a dynamically generated
supernovae wind model was coupled to models of galactic chemical evolution
with the aim of reducing a primordial D/H abundance of 2 $\times 10^{-4}$
to the present ISM value without overproducing heavy elements and 
remaining within the other observational constraints typically 
imposed on such models.
In Figure 13, the evolution of D/H and 
\he3/H is
shown as a function of time in several representative models
with significant deuterium destruction factors (see ref 45. for
details).  However,
as one can plainly see, \he3 is grossly overproduced (the deuterium data is
represented by squares and \he3 by circles). 

\begin{figure}
\hspace{0.5truecm}
\epsfysize=14truecm
\epsfbox{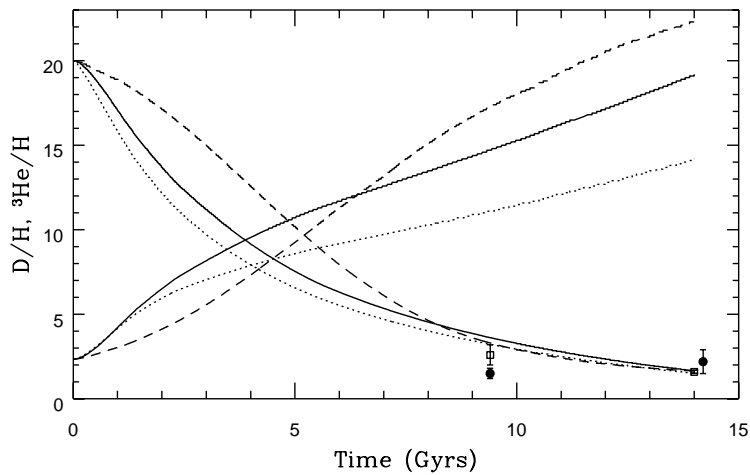}
\vspace{-7truecm}
\baselineskip=2ex
\caption { The evolution of D/H and \he3/H with time in units of $10^{-5}$.}
\end{figure}

\baselineskip=3ex
The overproduction of \he3 relative to the solar meteoritic value seems to be a
generic feature of chemical evolution models when \he3 production in low mass
stars is included. This result appears to be independent of the 
chemical evolution
model and is directly related to the assumed stellar yields of \he3.
It has recently been suggested that at least some low mass
stars may indeed be net destroyers of \he3 if one includes
the effects of rotational mixing in low mass stars on the
red giant branch \cite{char,bm}. The extra  
mixing does not take place for stars which do not undergo a helium core
flash (i.e. stars $>$ 1.7 - 2 M$_\odot$ ).  Thus stars with masses {\it less than}
1.7 M$_\odot$ are responsible for the \he3 destruction. 
Using the yields of Boothroyd and Malaney \cite{bm}, it was shown \cite{osst}
that these reduced \he3 yields in low mass stars can account for the
relatively low solar and present day \he3/H abundances observed.
In fact, in some cases, \he3 was underproduced.  To account for the \he3 evolution
and the fact that some low mass stars must be producers 
of \he3 as indicated by the
planetary nebulae data, it was suggested that the new yields apply
only to a fraction (albeit large) of low mass stars \cite{osst,gal}. 
The corresponding evolution \cite{osst} of 
D/H and \he3/H is shown in Figure 14.

\begin{figure}
\hspace{0.5truecm}
\epsfysize=14truecm\epsfbox{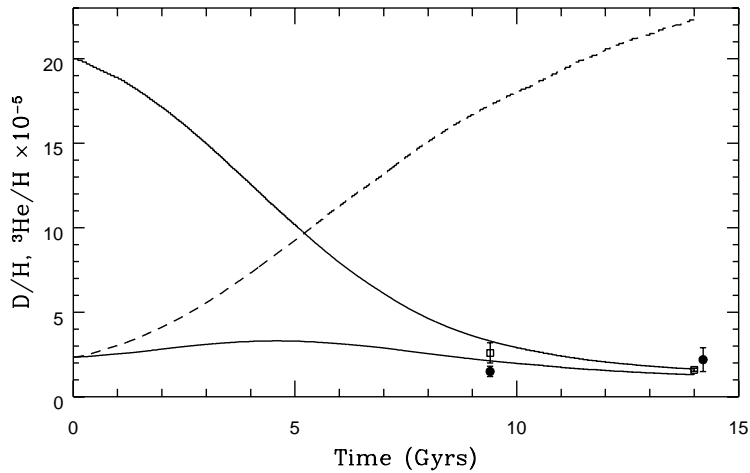}
\vspace{-7truecm}
\baselineskip=2ex
\caption { The evolution of D/H and \he3/H with time using the reduced
\he3 yields of ref. 47. The dashed curve is the same as in Figure 13, using
standard \he3 yields.}
\end{figure}

\subsection{Likelihood Analysis}

Monte Carlo techniques have proven to be a useful form of analysis for big
bang nucleosynthesis \cite{kr,kk,hata1}. An
analysis of this sort was performed \cite{fo}
using only \he4 and \li7. It should be noted that in principle, two
elements should be sufficient for not only constraining the one parameter 
($\eta$) theory of BBN, but also for testing for consistency. 
The procedure begins by establishing likelihood functions for the theory and
observations. For example, for \he4, the theoretical likelihood 
function takes the
form
\beq
L_{\rm BBN}(Y,Y_{\rm BBN}) 
  = e^{-\left(Y-Y_{\rm BBN}\left(\eta\right)\right)^2/2\sigma_1^2}
\label{gau}
\eeq
where $Y_{\rm BBN}(\eta)$ is the central value for the \he4 mass fraction
produced in the big bang as predicted by the theory at a given value of $\eta$,
and $\sigma_1$ is the uncertainty in that  value derived from the Monte Carlo
calculations \cite{hata1} and is a measure of the theoretical 
uncertainty in the
big bang calculation. Similarly one can write down an expression for the
observational likelihood function. Assuming Gaussian errors,
the likelihood function for the observations would
take a form similar to that in (\ref{gau}).

\begin{figure}
\hspace{0.5truecm}
\epsfysize=7truecm\epsfbox{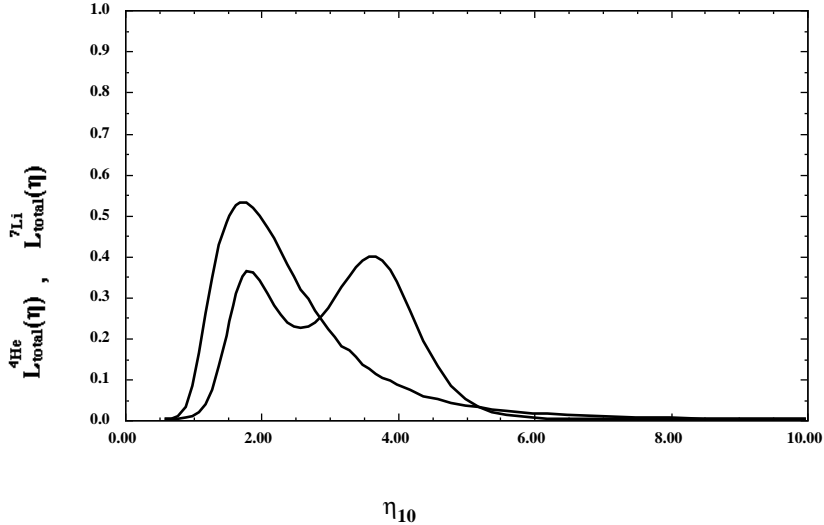}

\caption { \baselineskip=2ex Likelihood distribution for each of \he4 and \li7,
shown as a  function of $\eta$.  The one-peak structure of the \he4 curve
corresponds to its monotonic increase with $\eta$, while
the two peaks for \li7 arise from its passing through a minimum.}
\label{fig:fig1}
\end{figure}

A total likelihood 
function for each value of $\eta$ is derived by
convolving the theoretical
and observational distributions, which for \he4 is given by
\beq
{L^{^4{\rm He}}}_{\rm total}(\eta) = 
\int dY L_{\rm BBN}\left(Y,Y_{\rm BBN}\left(\eta\right)\right) 
L_{\rm O}(Y,Y_{\rm O})
\label{conv}
\eeq
An analogous calculation is performed \cite{fo} for \li7. 
The resulting likelihood
functions from the observed abundances given in Eqs. (\ref{he4}) 
  and (\ref{li})
is shown in Figure \ref{fig:fig1}. As one can see 
there is very good agreement between \he4 and \li7 in the vicinity
of $\eta_{10} \simeq 1.8$. The double peaked nature of the \li7
likelihood function is due to the presence of a minimum in the
 predicted lithium abundance.  For a given observed value of \li7, there
are two likely values of $\eta$.

\begin{figure}
\hspace{0.5truecm}
\epsfysize=7truecm\epsfbox{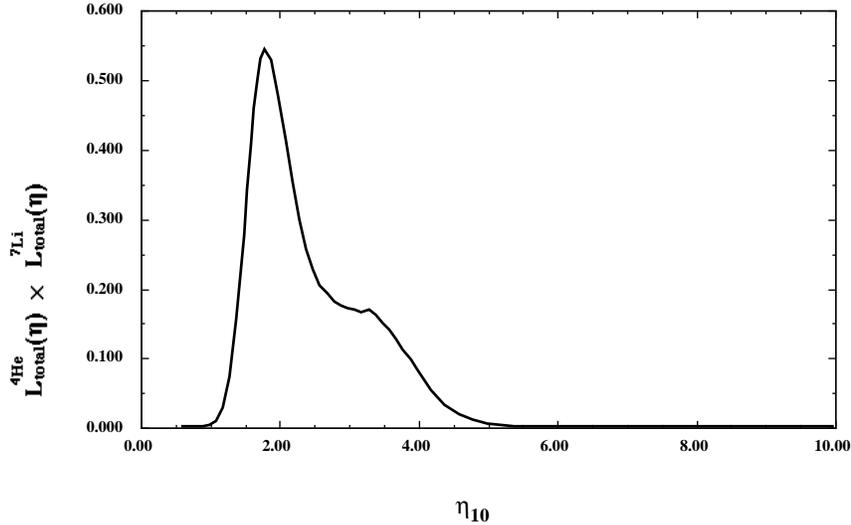}
\baselineskip=2ex
\caption { Combined likelihood for simultaneously fitting \he4 and \li7,
as a function of $\eta$.}
\label{fig:fig2}
\end{figure}

\baselineskip=3ex

The combined likelihood, for fitting both elements simultaneously,
is given by the product of the two functions in Figure \ref{fig:fig1}
and is shown  in Figure \ref{fig:fig2}.
{}From Figure \ref{fig:fig1} it is clear that \he4 overlaps
the lower (in $\eta$) \li7 peak, and so one expects that 
there will be concordance
in an allowed range of $\eta$ given by the overlap region.  
This is what one finds in Figure \ref{fig:fig2}, which does
show concordance and gives a preferred value for $\eta$, 
$\eta_{10}  = 1.8^{+1.2}_{-.2}$ corresponding to 
\beq
\Omega_B h^2 = .006^{+.004}_{-.001}
\label{omega}
\eeq 

Thus,  we can conclude that 
the abundances of 
\he4 and \li7 are consistent, and select an $\eta_{10}$ range which
overlaps with (at the 95\% CL) the longstanding favorite
 range around $\eta_{10} = 3$.
Furthermore, by finding concordance  
using only \he4 and \li7, we deduce that
if there is problem with BBN, it must arise from 
D and \he3 and is thus tied to chemical evolution or the stellar evolution of
\he3. The most model-independent conclusion is that standard
BBN  with $N_\nu = 3$ is not in jeopardy.
 It is interesting to
note that the central (and strongly)  peaked
value of $\eta_{10}$ determined from the combined \he4 and\li7 likelihoods
is at $\eta_{10} = 1.8$.  The corresponding value of D/H is 1.8 $\times 
10^{-4}$, very close \cite{dar} to the high value  of D/H in quasar absorbers
\cite{quas1,quas3}. This is one of the main motivations for studying
galactic chemical evolution models with high initial D/H and strong 
D destruction histories.  Correspondingly, the the use of reduced 
\he3  yields is essential as described in the previous section.
Since  D and \he3 are monotonic functions of $\eta$, a prediction for 
$\eta$, based on \he4 and \li7, can be turned into a prediction for
D and \he3.  
 The corresponding 95\% CL ranges are D/H  $= (4.7 - 28)  \times
10^{-5}$ and \he3/H $= (1.3 - 2.7)  \times 10^{-5}$.

If we did have full confidence in the measured value of D/H in 
quasar absorption
systems, then we could perform the same statistical analysis 
using \he4, \li7, and
D. To include D/H, one would
proceed in much the same way as with the other two light elements.  We
compute likelihood functions for the BBN predictions as in
Eq. (\ref{gau}) and the likelihood function for the observations using
D/H = $(1.9 \pm 0.4) \times 10^{-4}$ i.e.,  using only the high
 value of D/H
here. These are
then convolved as in Eq.  (\ref{conv}).  
In Figure 17, the resulting normalized
distribution, $L^{{\rm D}}_{\rm total}(\eta)$ is super-imposed on
distributions appearing in Figure \ref{fig:fig1}. 
It is indeed startling how the three peaks, for
D, \he4 and \li7 are literally on top of each other.  In Figure 18, 
the combined distribution is shown.
We now  have a very clean distribution and prediction 
for $\eta$, $\eta_{10}  = 1.75^{+.4}_{-.1}$ corresponding to $\Omega_B h^2 =
.006^{+.001}_{-.0004}$,
with the peak of the distribution at $\eta_{10} = 1.75$.  
The absence of any overlap with the high-$\eta$ peak of the \li7
distribution has considerably lowered the upper limit to $\eta$. 
Overall, the concordance limits in this case are dominated by the 
deuterium likelihood function.

\begin{figure}
\hspace{0.5truecm}
\epsfysize=7truecm\epsfbox{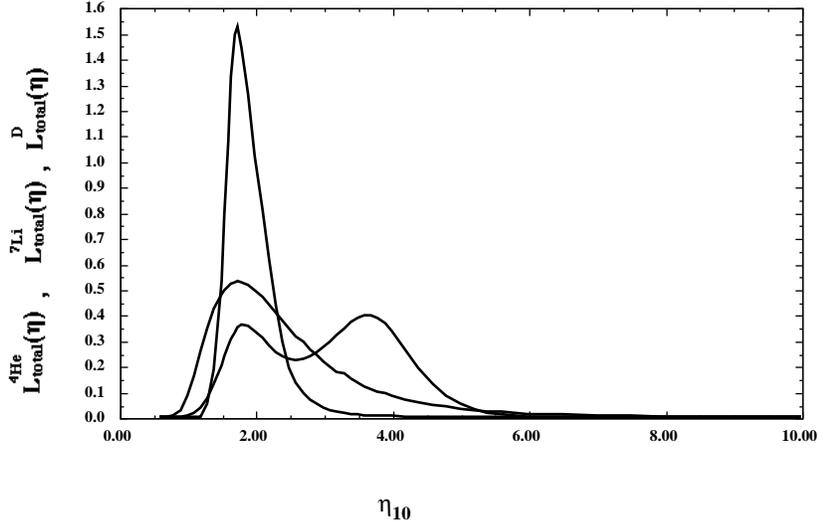}
\baselineskip=2ex
\caption{ As in Figure 16, with the addition of the likelihood 
distribution for D/H.}
\label{fig:fig4}
\end{figure}

\baselineskip=3ex

\begin{figure}
\hspace{0.5truecm}
\epsfysize=7truecm\epsfbox{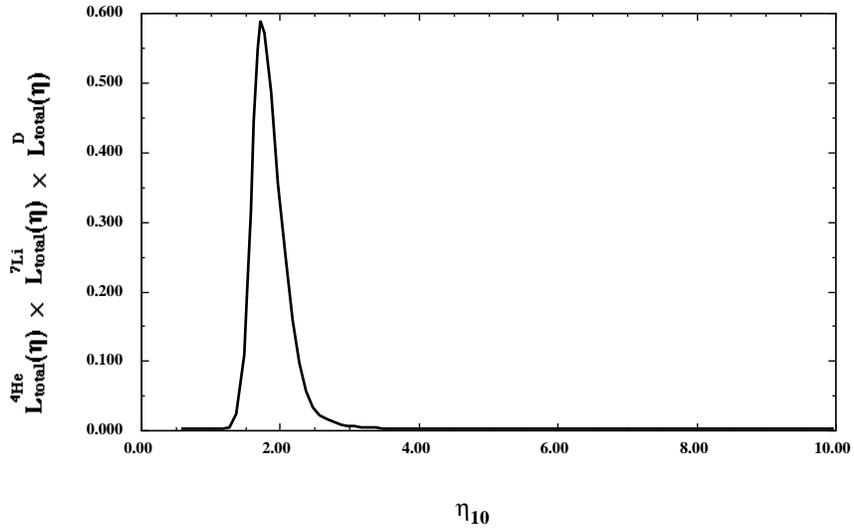}
\baselineskip=2ex
\caption { Combined likelihood for simultaneously fitting 
\he4 and \li7, and D 
as a function of $\eta$.
}
\label{fig:fig5}
\end{figure}

\baselineskip=3ex

For the most part I have concentrated on the high D/H measurements in the
likelihood analysis. If instead, we assume that the low value \cite{quas2}
of D/H = $(2.5 \pm 0.5) \times 10^{-5}$ is the primordial abundance,
then we can again compare the likelihood distributions as in Figure 17,
now substituting the low D/H value. As one can see from Figure 19, there 
is now hardly any overlap between the D and the \li7 distributions
and essentially no overlap with the \he4 distribution.  The combined distribution
shown in Figure 20 is compared with that in Figure 18.
Though one can not use this likelihood analysis to prove
 the correctness of the high
D/H measurements or the incorrectness of the low D/H measurements,
the analysis clearly shows the difference in compatibility between the
two values of D/H and the observational determinations of \he4 and \li7.
To {\em make} the low D/H measurement compatible, one would have to argue
for a shift upwards in \he4 to a primordial value of 0.249 (a shift by 0.015)
which is certainly not warranted
at this time by the data, and a \li7 depletion factor of 
about 3, which exceeds recent upper limits to the amount of depletion
\cite{cv}.

\begin{figure}
\hspace{0.5truecm}
\epsfysize=7truecm\epsfbox{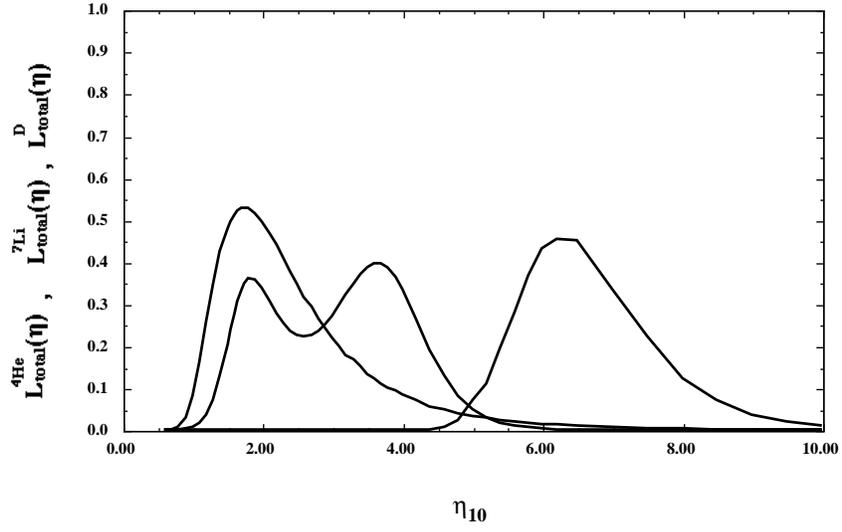}
\baselineskip=2ex
\caption { As in Figure 17, with the likelihood 
distribution for low 
D/H. }
\label{fig:fig6}
\end{figure}

\baselineskip=3ex

\begin{figure}
\hspace{0.5truecm}
\epsfysize=7truecm\epsfbox{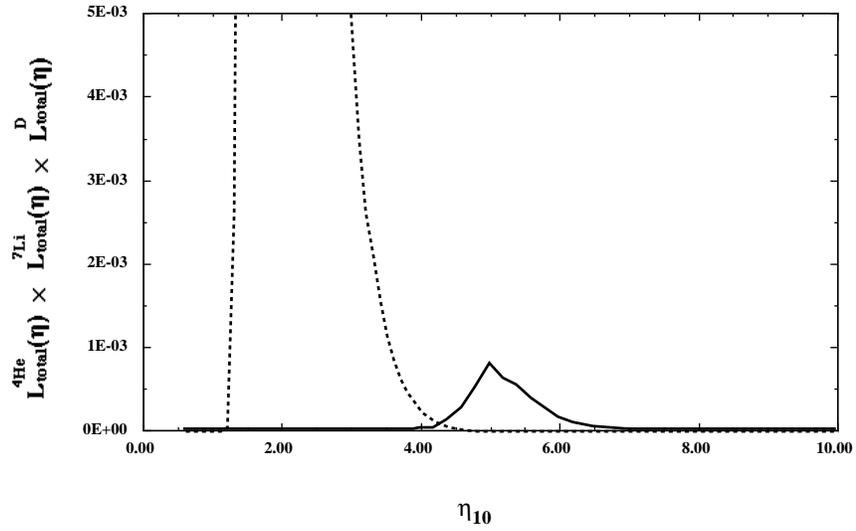}
\baselineskip=2ex
\caption { Combined likelihood for simultaneously fitting 
\he4 and \li7, and  low D/H
as a function of $\eta$. The dashed curve represents the combined distribution
shown in Figure 18.
}
\label{fig:fig7}
\end{figure}

\baselineskip=3ex

The implications of the resulting predictions from big bang nucleosynthesis
on dark matter are clear.  First, if $\Omega = 1$ (as predicted by 
inflation), and $\Omega_B \la 0.1$ which is certainly a robust
conclusion based on D/H, then non-baryonic dark matter is a necessity.
Second, on the scale of small groups of galaxies, $\Omega \ga 0.05$,
and is expected to sample the dark matter in galactic halos.
This value is probably larger than the best estimate for $\Omega_B$
from equation (\ref{omega}). $\Omega_B h^2 = 0.0065$ corresponds to 
$\Omega_B = 0.025$ for $h = 1/2$.  In this event,
some non-baryonic dark matter in galactic halos is required. 
This conclusion is unchanged
by the inclusion of the high D/H measurements in QSO absorbers.
In contrast \cite{2d}, the low D/H measurements would imply that
$\Omega_B h^2 = 0.023$ allowing for the possibility that $\Omega_B
\simeq 0.1$.  In this case, no non-baryonic dark matter is
required in galactic halos.  However, I remind the reader that
the low D/H is at present not consistent with either the observations of
\he4 nor \li7 and their interpretations as being primordial abundances.
I will return to the subject of dark matter in the next lecture.

\subsection{Constraints from BBN}

Limits on particle physics beyond the standard model are mostly sensitive to
the bounds imposed on the \he4 abundance. 
As is well known, the $^4$He abundance
is predominantly determined by the neutron-to-proton ratio just prior to
nucleosynthesis and is easily estimated assuming that all neutrons are
incorporated into \he4 (see Eq. (\ref{ynp})).
As discussed earlier, the neutron-to-proton
ratio is fixed by its equilibrium value at the freeze-out of 
the weak interaction rates at a temperature $T_f \sim 1$ MeV modulo the
occasional free neutron decay.  Furthermore, freeze-out is determined by the
competition between the weak interaction rates and the expansion rate of the
Universe
\begin{equation}
{G_F}^2 {T_f}^5 \sim \Gamma_{\rm wk}(T_f) = H(T_f) \sim \sqrt{G_N N} {T_f}^2
\label{comp} \label{freeze}
\end{equation}
where $N$ counts the total (equivalent) number of relativistic particle
species. The presence
of additional neutrino flavors (or any other relativistic species) at 
the time of nucleosynthesis increases the overall energy density
of the Universe and hence the expansion rate leading to a larger 
value of $T_f$, $(n/p)$, and ultimately $Y_p$.  Because of the
form of Eq. (\ref{comp}) it is clear that just as one can place limits
\cite{ssg} on $N$, any changes in the weak or gravitational coupling constants
can be similarly constrained (for a recent discussion see ref. 56).
In concluding this lecture, I will discuss the current constraint on
$N_\nu$ the number of particle species (in neutrino units) and the
limit on the strength of new interactions, if 3 right-handed (nearly massless)
neutrinos are assumed to exist.

In the past, \he3 (together with D) has stood out 
in its importance for BBN, because 
it  provided a (relatively large) lower limit for the baryon-to-photon
ratio \cite{ytsso}, $\eta_{10} > 2.8$. This limit for a long 
time was seen to be
essential because it provided the only means for bounding $\eta$ from below
and in effect allows one to set an upper limit on the number of neutrino
flavors \cite{ssg}, $N_\nu$, as well as other constraints on particle physics
properties. That is, the upper bound to $N_\nu$ 
is strongly dependent on the lower bound to
$\eta$.  This is easy to see: for lower $\eta$, the \he4 abundance drops,
allowing for a larger $N_\nu$, which would raise the \he4 abundance.
However, for $\eta < 4 \times 10^{-11}$, corresponding to $\Omega_B h^2 \la
.001-.002$, which is not too different from galactic mass densities, 
there is no
bound whatsoever on $N_\nu$ \cite{ossty}. 
Of course, with the improved data on
\li7, we do have lower bounds on $\eta$ which exceed $10^{-10}$.

Because, of new observations of D and \he3, and the new theoretical
work on chemical evolution sparked by these observations, 
the bound on $N_\nu$ which is tied directly 
to these isotopes, should be called into question.
As described earlier, the limits due to \he3 are ultimately
tied to the assumed yields of low mass stars.
Using the reduced yields as depicted in Figure 14, consistent values of 
$\eta < 2.8$ are certainly possible.    Ultimately, as I have said
repeatedly, D/H measurements in quasar absorption systems may soon
resolve this issue.  However, the lower values of $\eta$, 
relax the bounds on the number of neutrino flavors.

As discussed above, the limit on $N_\nu$ comes about via the 
change in the expansion rate given by the Hubble parameter,
\beq
H^2 = {8 \pi G \over 3} \rho = {8 \pi^3  G \over 90} [N_{\rm SM} 
+ {7 \over 8} \Delta N_\nu] T^4
\eeq
when compared to the weak interaction rates. Here $N_{\rm SM}$
refers to the standard model value for N. At $T \sim 1$ MeV,
$N_{\rm SM} = 43/4$. Additional degrees of freedom will 
lead to an increase in the freeze-out temperature eventually leading to
a higher \he4 abundance. In fact, one 
can parameterize the dependence of $Y$ on $N_\nu$ by 
\beq
Y = 0.2262 + 0.0131 (N_\nu - 3) + 0.0135 \ln \eta_{10} 
\label{YY}
\eeq
in vicinity of $\eta_{10} \sim 2$.  Eq. (\ref{YY}) also shows
the weak (log) dependence on $\eta$. However, rather than use
(\ref{YY}) to obtain a limit, it is preferable to use 
the likelihood method.

 Just as \he4 and \li7 were sufficient to
determine a value for $\eta$,  a limit on $N_\nu$ can be obtained
as well \cite{fo,oth3}. The likelihood approach
utilized above can be extended to include $N_\nu$ as a free parameter.
Since the light element abundances can be computed as functions
of both $\eta$ and $N_\nu$,  the
likelihood function can be defined by \cite{oth3}
\beq
L_{\rm BBN}(Y,Y_{\rm BBN}) 
  = e^{-\left(Y-Y_{\rm BBN}\left(\eta,N_\nu\right)\right)^2/2\sigma_1^2}
\label{gau1}
\eeq
and \beq
{L^{^4{\rm He}}}_{\rm total}(\eta,N_\nu) = 
\int dY L_{\rm BBN}\left(Y,Y_{\rm BBN}\left(\eta,N_\nu\right)\right) 
L_{\rm O}(Y,Y_{\rm O})
\label{conv1}
\eeq
Again, similar expressions are needed for \li7 and D. 
A three-dimensional view of the combined likelihood functions \cite{oth3} 
is shown in Figure 21.
In this case the high and low $\eta$ maxima of Figure 15, show up
as peaks in the $L-\eta-N_\nu$ space
($L_{47}$ when D/H is neglected and $L_{247}$ when D/H is included).
The peaks of the distribution as well as the
allowed ranges of $\eta$ and $N_\nu$ are  
\begin{figure}
\hspace{0.5truecm}
\epsfysize=15truecm\epsfbox{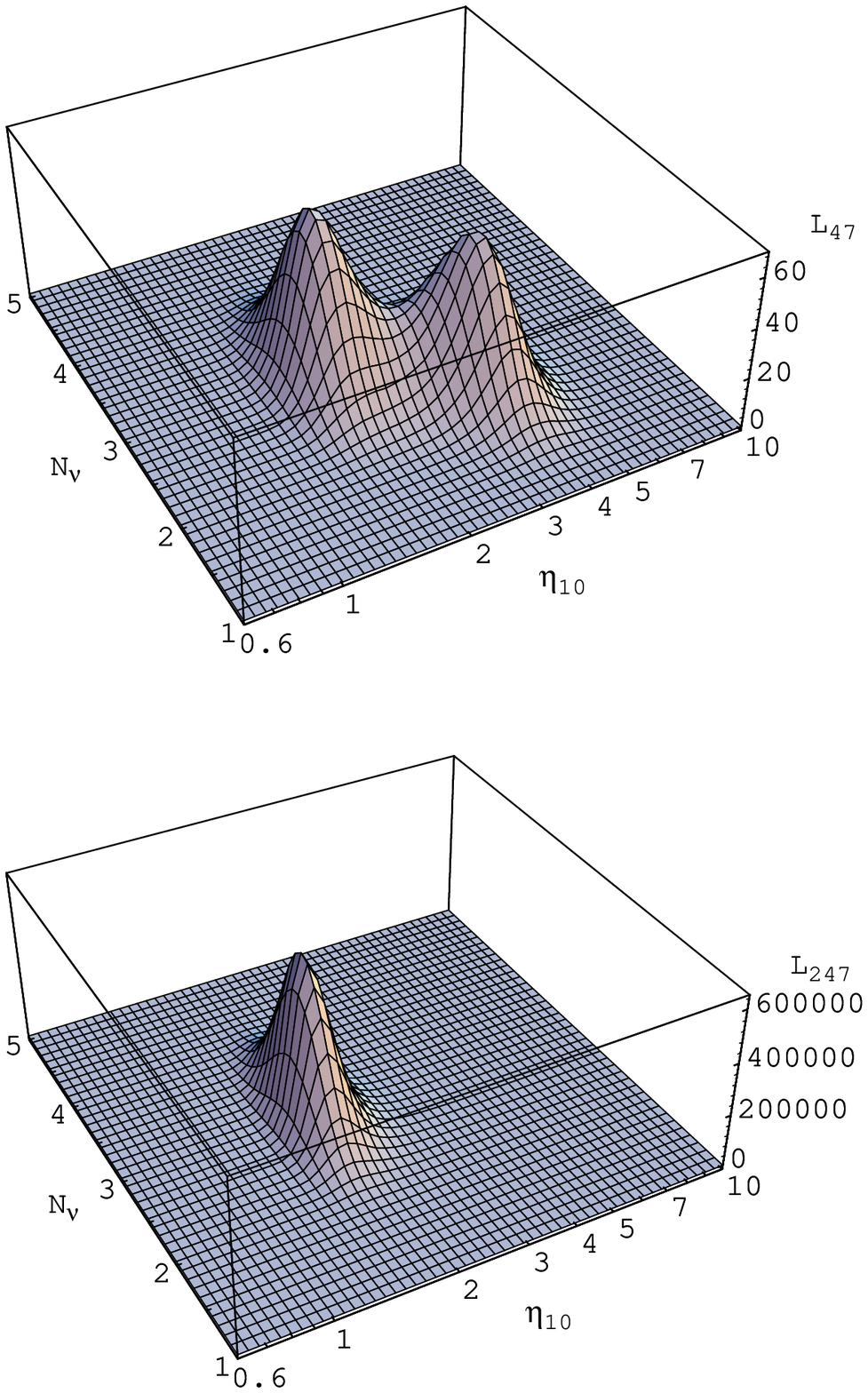}
\baselineskip=2ex
\caption {The combined two-dimensional 
likelihood functions for simultaneously fitting 
\he4 and \li7 in the top panel, and including D in the lower one
as functions of both $\eta$ and $N_\nu$.
}
\label{fig:fig1ai}
\end{figure}
\begin{figure}
\hspace{0.5truecm}
\epsfysize=15truecm\epsfbox{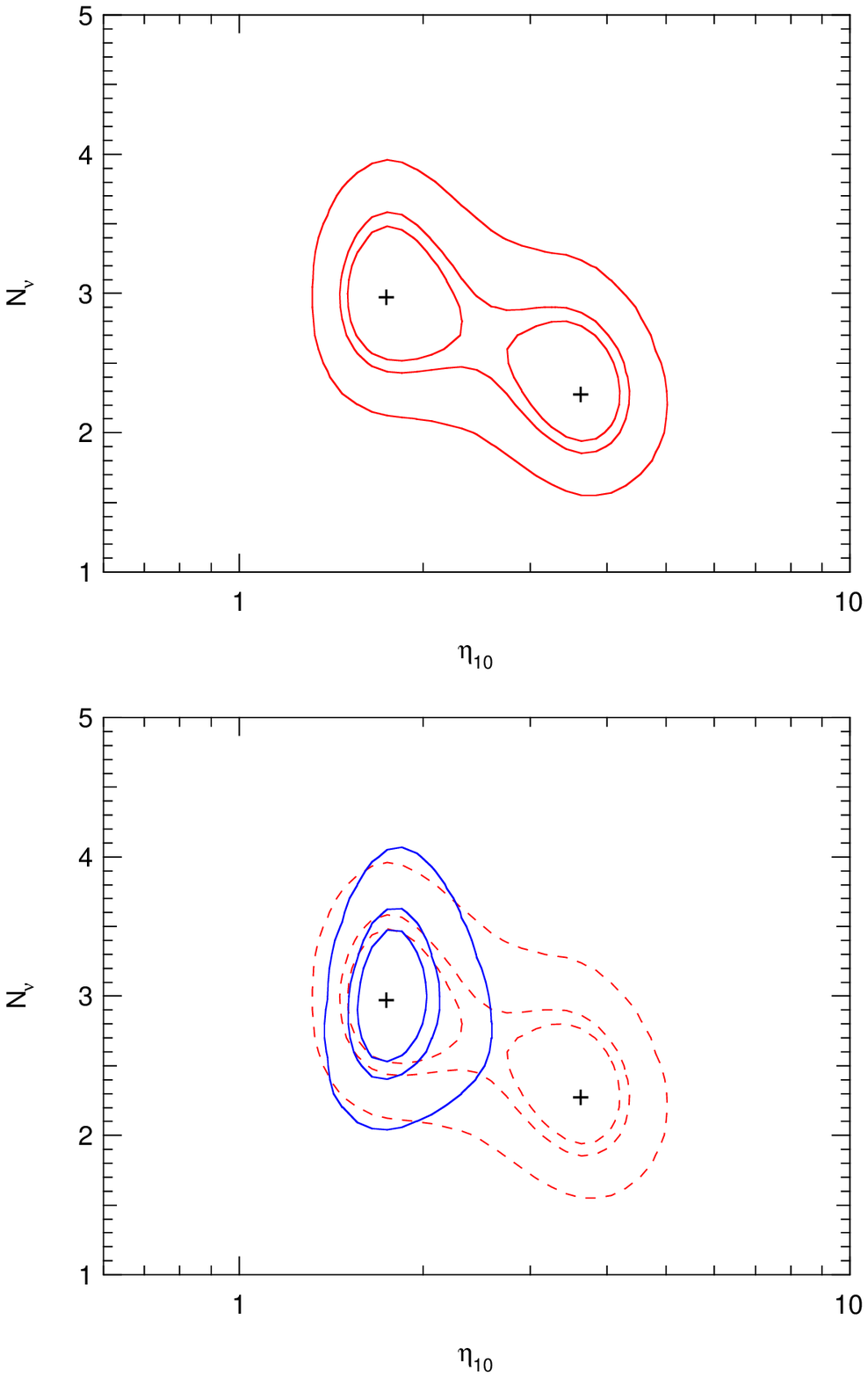}
\caption{The top panel shows contours in the combined
		likelihood function for \he4 and \li7.
		The contours represent 50\% (innermost), 68\% and 95\% 
		(outermost) confidence levels.  The crosses mark the points 
		of maximum likelihood.  Also shown is the 
		equivalent result when D is included.}
\baselineskip=2ex
\label{fig:fig2ai}
\end{figure}
more easily discerned in the 
contour plot of Figure 22 which shows the 50\%,
68\% and 95\% confidence level contours in the two likelihood functions.  
The crosses show the location of the 
peaks of the likelihood functions.
$L_{47}$ peaks at $N_\nu=3.0$, $\eta_{10}=1.8$ (in agreement with 
our previous results \cite{fo}) and at $N_\nu=2.3$,
$\eta_{10}=3.6$.  The 95\% confidence level allows the following ranges
in $\eta$ and $N_\nu$
\beq
1.6\le N_\nu\le4.0  \qquad \qquad
1.3\le\eta_{10}\le 5.0 
\eeq
Note however that the ranges in $\eta$ and $N_\nu$ are strongly
correlated as is evident in Figure 22.
Since the deuterium likelihood function picks out a small range of values 
of $\eta$, largely independent of $N_\nu$, its effect on $L_{247}$ is 
to eliminate one of the two peaks in $L_{47}$. $L_{247}$ also 
peaks at $N_\nu=3.0$, $\eta_{10}=1.8$. 
In this case
the 95\% contour gives the ranges
\beq
2.0\le N_\nu\le4.1 \qquad \qquad
1.4\le\eta_{10}\le 2.6 
\eeq

One should recall that the limit derived above is not meant for neutrinos
in the strictest sense.  That is, the limit is only useful
when applied to additional particle degrees of freedom which 
necessarily do not couple to the Z$^o$. For very weakly interacting 
particles, one must take into account the reduced 
abundance of these particles at the time of nucleosynthesis\cite{oss}.  
As discussed in the first lecture, the number of neutrinos today is reduced
relative to the number of photons by
$(T_\nu/T_\gamma)^3  = 4/11$.  
For some new particle, $\chi$, which decoupled at $T_d > 1$ MeV, 
the same argument based on the conservation of entropy tells us that
\beq
({T_\chi \over T_\gamma})^3 = {43 \over 4 N(T_d)}
\label{decx}
\eeq
Thus we can translate the bound on $N_\nu$, which is really a bound 
on the additional energy density at nucleosynthesis
\beq
\Delta \rho = {\pi^2 \over 30} \left[ \sum g_B ({T_B \over T})^4
 + {7 \over 8} \sum g_F ({T_F \over T})^4 \right] T^4
\eeq
for additional boson states with $g_B$ degrees of freedom and
fermion states with $g_F$ degrees of freedom.
At nucleosynthesis $T = T_\nu = T_\gamma$ and the limit $N_\nu < 4$ 
becomes
\beq 
{8 \over 7} \sum {g_B \over 2} ({T_B \over T_\nu})^4
 +  \sum {g_F \over 2} ({T_F \over T})^4 < 1
\eeq
Such a limit would allow a single additional scalar
degree of freedom (which counts as ${4 \over 7}$) such as a majoron. 
On the other hand, 
in models with right-handed interactions, 
and three right-handed neutrinos, the
constraint is severe. 
The right-handed states must have decoupled early
enough to ensure $(T_{\nu_R}/T_{\nu_L})^4 < 1/3$. The
temperature of a decoupled state is easily determined from (\ref{decx}). 
 Three right-handed neutrinos would require  $N(T_d) \ga 25$, 
which from Figure 1 implies that $T_d
> 140$ MeV, conservatively assuming a QCD transition temperature
of 150 MeV. If right-handed interactions are mediated by additional gauge
interactions, associated with some scale $M_{Z'}$,
and if we assume that the right handed interactions scale
as $M_{Z'}^4$ with a standard model-like coupling, 
then the decoupling temperature of the right handed interactions
is related to $M_{Z'}$
by 
\beq
({{T_d}_R \over {T_d}_L})^3 \sim ({M_{Z'} \over M_Z})^4
\eeq
which for ${T_d}_L \sim 3$ MeV ( a more accurate value that the 
1 MeV estimate) and ${T_d}_L \ga 140$ MeV,
we find that the associated mass scale becomes $M_{Z'} \ga 1.6$ TeV!
In general this constraint is very sensitive to the BBN limit on $N_\nu$.
In Figure 23, the allowed number of neutrino degrees of freedom are
shown as a function of their decoupling temperature for the case
of $N_\nu < 4$ and $N_\nu < 3.3$, shown for comparison.

\begin{figure}
\hspace{0.5truecm}
\epsfysize=7truecm\epsfbox{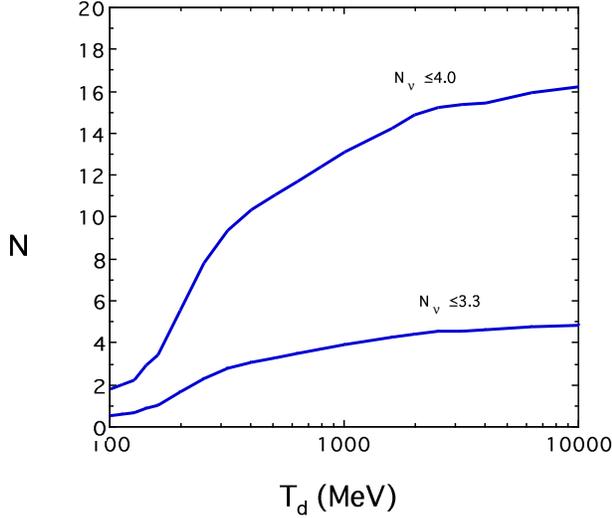}
\baselineskip=2ex
\caption { Limits on neutrino-like degrees of freedom.}
\end{figure}

\subsection{Summary} 

To summarize on the subject of big bang nucleosynthesis, 
I would assert that one
can conclude that the present data on the abundances of the light element
isotopes are consistent with the standard model of big bang 
nucleosynthesis. Using
the the isotopes with the best data, \he4 and
\li7, it is possible to constrain the theory and obtain a best value for the
baryon-to-photon ratio of $\eta_{10} = 1.8$, a corresponding 
value $\Omega_B h^2 =
.0065$ and
\beq
\begin{array}{ccccc}
1.4 & < & \eta_{10} & < &  4.2  \qquad 95\% {\rm CL} \nonumber \\
.005 & < & \Omega_B h^2 & < & .015  \qquad 95\% {\rm CL}
\label{res2}
\end{array}
\eeq
For $0.4 < h < 1$, we have a range $ .005 < \Omega_B < .10$.
This is a rather low value for the baryon density
 and would suggest that much of the galactic dark matter is
non-baryonic \cite{vc}. These predictions are in addition 
consistent with recent
observations of D/H in quasar absorption systems which show a high value.
Difficulty remains however, in matching the solar \he3 abundance, suggesting a
problem with our current understanding of galactic chemical evolution or the
stellar evolution of low mass stars as they pertain to \he3.

\section{Dark Matter}

 There is increasing evidence  
that relative to the visible matter in the Universe, which 
is in the form of baryons, there is considerably more
 matter in the Universe 
that we don't see \cite{dm}.
	Here, I will review some of the motivations for the existence of dark
 matter in the Universe.  The best observational evidence 
is found on the scale of galactic halos and comes from the 
observed flat rotation curves of galaxies. There is also mounting
evidence for dark matter in elliptical galaxies, as well as clusters
of galaxies coming from X-ray observations of these objects.
Also, direct evidence has been obtained through the study of
gravitational lenses.
In theory, we 
believe there is much more matter because 1) inflation 
tells us so (and there is at present no good alternative to inflation)
 and 2) our current understanding of galaxy 
formation only makes sense if there is more matter than we see.  
	One can also make a strong case for the existence of non-baryonic 
dark matter in
particular. The recurrent problem  with baryonic dark matter
is that not only is it very difficult to hide baryons, but 
the standard model of primordial nucleosynthesis
would have to be discarded
if all of the dark matter is baryonic. 
 Fortunately, as will be covered at length
in these proceedings, there are several attractive alternatives to baryonic
dark matter.

There is, in fact, good evidence for dark matter 
on the scale of galaxies (and their halos). 
 Assuming that galaxies are in virial equilibrium,
 one expects that by Newton's Laws one can relate 
the mass at a given distance $r$, from the center of 
a galaxy to its rotational velocity
\beq
	M(r) \propto v^2 r/G_N 	
\eeq
The rotational velocity, $v$, is measured \cite{fg,bos,rft}
 by observing 21 cm 
emission lines in HI regions (neutral hydrogen) beyond the point 
where most of the light in the galaxy ceases.  A compilation 
of nearly 1000
 rotation curves of spiral galaxies has been plotted in ref.
66 as a function
of $r$ for varying brightnesses.  If the bulk of the mass is 
associated with light, then beyond the point where most of the light 
stops $M$ would be 
constant and $v^2  \propto 1/r$.  This is not the case, as
the rotation
 curves appear to be flat, i.e., $v \sim$ constant outside the
 core of the galaxy. This implies that $M \propto r$ beyond the point
 where the light stops.  This is one of the strongest pieces of 
evidence for the existence of dark matter. Velocity measurements indicate
dark matter in elliptical galaxies as well \cite{sag}.

Galactic rotation curves are not the only observational indication for the
existence of dark matter.  X-ray emitting hot gas in elliptical galaxies 
also provides an important piece of evidence for dark matter. 
 As an example, 
consider the large elliptical M87.  The detailed profiles of the temperature
and density of the hot X-ray emitting gas have been mapped out \cite{fgo}.
By assuming hydrostatic equilibrium, these measurements allow 
one to determine 
the overall mass distribution in the galaxy necessary to bind the hot gas.
Based on an isothermal model with temperature $kT = 3$keV (which leads
to a conservative estimate of the total mass), Fabricant and
 Gorenstein \cite{fgo}
predicted that the total mass out to a radial distance
 of 392 Kpc is $5.7 \times 10^{13} M_\odot$,
whereas the mass in the hot gas is only $2.8 \times 10^{12} 
M_\odot$ or only 5\%
of the total. The visible mass is expected to contribute 
only 1\% of the total.
The inferred value of $\Omega$ based on M87 would be $\sim 0.2$.

M87 is not the only example of an elliptical galaxy in which X-ray emitting
hot gas is observed to indicate the presence of dark matter.  
In addition,  similar inferences regarding the
existence of dark matter can be made from the X-ray emission 
from small groups of
galaxies \cite{mush,F2}.

On very large scales, it is possible to get an estimate of $\Omega$ from the
distribution of peculiar velocities. On scales, $\lambda$, 
where perturbations, $\delta$, are still
small, peculiar velocities can be expressed \cite{peeb}
 as $v \sim H \lambda \delta \Omega^{0.6}$.
 On these scales, older measurements of the peculiar velocity field from 
the IRAS galaxy catalogue indicate that indeed $\Omega$ 
is close to unity \cite{iras}. Some of the new data \cite{newiras} 
indicates perhaps a lower value in range 0.2 -- 0.5, but does 
not conclusively exclude $\Omega = 1$.
Another piece of evidence on large scales is available from 
gravitational lensing
\cite{tyson}. The systematic lensing of the roughly 150,000 
galaxies per deg$^2$
at redshifts between $z = 1 - 3$ into arcs and arclets allow 
one to trace the
matter distribution in a foreground cluster. 
 Unfortunately, none of these 
observations reveal the identity of the dark matter.

Theoretically, there is no lack of support for the dark matter hypothesis.
The standard big bang model including inflation almost requires that
$\Omega = 1$ \cite{infl}.
	If this is the case and $\Omega = 1$, then we know two things:  Dark
matter exists, since we don't see $\Omega = 1$ in luminous objects, and most
(at least 90\%) of the dark matter is not baryonic.  The latter
conclusion is a result of our previous discussion on BBN which
constrains the baryon-to-photon ratio and hence $\Omega_B$ as in eq.
(\ref{res2}).  Thus $1-\Omega_B$ is not only dark but
also non-baryonic. 

	Another important piece of theoretical evidence for dark
 matter comes from the simple fact that we are here living in a galaxy.
The type of perturbations produced
 by inflation \cite{press} are, in most models,
 adiabatic perturbations ($\delta\rho/\rho \propto
 \delta T/T)$, and I
 will restrict my attention to these.  Indeed, the perturbations
produced by inflation
 also have the very nearly scale-free spectrum described by 
Harrison and Zeldovich \cite{hz}.  When produced, scale-free perturbations 
fall off as $\frac{\delta \rho}{\rho} \propto l^{-2}$
 (increase as the square of the wave number). 
 At early times $\delta\rho/\rho$ grows as $t$
 until the time when the horizon scale (which is
 proportional to the age of the Universe) is comparable to $l$.  At later 
times, the growth halts (the mass contained within the volume $l^3$  
has become smaller than the Jean's mass) and   
  $\frac{\delta \rho}{\rho} = \delta$ (roughly) independent of the scale $l$.
 When the Universe becomes matter dominated, the Jean's mass 
drops dramatically and growth continues as $\frac{\delta \rho}{\rho} \propto
 R \sim 1/T$. 

 The transition to matter 
dominance
is determined by setting the energy densities in radiation
(photons and any massless  neutrinos) equal to the energy density in  matter
(baryons and any dark matter).  For three massless  neutrinos and baryons (no
dark matter), matter dominance begins at
\beq
	T_m  = 0.22 m_B \eta	
\eeq
and for $\eta < 3.8 \times 10^{-10}$, this corresponds to
$T_m < 0.08$ eV.

	Because we are considering adiabatic perturbations,
 there will be anisotropies produced in the microwave 
background radiation on the order of $\delta T/T \sim \delta$.  
The value of $\delta$, the amplitude of the density fluctuations at horizon 
crossing, has now been determined by COBE \cite{cobe}, $\delta =
(5.7 \pm 0.4) \times 10^{-6}$.  Without the existence of dark matter,
 $\delta \rho/\rho$ in baryons could then achieve a maximum value of only
$\delta\rho/\rho \sim A_\lambda \delta(T_m/T_o)  
\la 2 \times 10^{-3}A_\lambda$,
where $T_o = 2.35 \times 10^{-4}$ eV is the present temperature of the
microwave background and $A_\lambda
\sim 1-10$ is a scale dependent growth factor. 
 The overall growth in $\delta \rho / \rho$ is too small to argue
 that growth has entered a nonlinear regime needed to explain
 the large value ($10^5$) of $\delta\rho/\rho$ in galaxies.

	Dark matter easily remedies this dilemma in the following way.
 The transition to matter dominance is determined by setting equal 
to each other the energy densities in radiation (photons and any massless 
neutrinos) and matter (baryons and any dark matter). 
While if we suppose that there exists 
dark matter with an abundance $Y_\chi = n_\chi/n_\gamma$  
(the ratio of the number density of $\chi$'s to photons) then
\beq
	T_m  = 0.22 m_\chi Y_\chi	
\eeq
Since we can write $m_\chi Y_\chi/{\rm GeV} = \Omega_\chi
 h^2/(4 \times 10^7)$,
we have $T_m/T_o = 2.4 \times 10^4 \Omega_\chi h^2$ which is 
to be compared with
350 in the case of baryons alone.  
The baryons, although still bound to the radiation until 
decoupling,  now see deep potential wells formed by the dark matter
 perturbations to fall into and are no longer required to 
grow at the rate $\delta \rho/\rho \propto R$.

	With regard to dark matter and galaxy formation, all forms 
of dark matter are not equal.  They can be distinguished 
by their relative temperature at $T_m$ \cite{bond}. Particles which are still 
largely relativistic at $T_m$ (like neutrinos or other particles with 
$m_\chi < 100$ eV) have the property \cite{free} that 
(due to free streaming) they
erase perturbations 
 out to very large scales given by the Jean's mass
\beq
	M_J  = 3 \times 10^{18}  {M_\odot \over {m_\chi}^2(eV)}	
\label{mj}
\eeq
Thus, very large scale structures form first and galaxies
 are expected to fragment out later.  Particles with this 
property are termed hot dark matter particles.  
Cold particles ($m_\chi > 1$ MeV) have the opposite behavior. 
 Small scale structure forms first aggregating to form 
larger structures later.  Neither candidate is completely 
satisfactory when the resulting structure is compared to the 
observations.  For more details, I refer the reader to reviews 
in refs. 62.

	Accepting the dark matter hypothesis, the first choice for a
 candidate should be something we know to exist, baryons.  
Though baryonic dark matter can not be the whole story if $\Omega = 1$, 
 the identity of the
 dark matter in galactic halos, which appear to contribute at the 
level of $\Omega \sim 0.05$,  remains an important question needing to be
resolved.  A baryon density of this magnitude is not excluded by 
nucleosynthesis. 
 Indeed we know some of the baryons are dark since $\Omega \la 0.01$ 
in the disk of the galaxy.

It is quite difficult, however, to hide large amounts 
of baryonic matter \cite{hio12}. Sites for halo baryons that 
have been discussed
include snowballs, which tend to sublimate, cold hydrogen gas, which 
must be supported against collapse, and hot gas, which can be excluded by the 
X-ray background. Stellar objects (collectively called MACHOs for macroscopic
compact halo objects) must either be so
small ( M $< 0.08$ M$_\odot$) so as not to have begun nuclear burning or so
massive so as to have undergone total gravitational collapse without the 
ejection of heavy elements.  Intermediate mass stars are generally quite
problematic because either they are expected to still reside on the 
main-sequence
today and hence would be visible, or they would have produced an 
excess of heavy
elements.

On the other hand, MACHOs are a candidate which are testable by the
gravitational microlensing of
stars in a neighboring galaxy such as the LMC \cite{pac}. By observing
millions of stars and examining their intensity as a function of time,
it may be possible to determine the presence of dark objects in our halo.
It is expected that during a lensing event, a star in the LMC will have
its  intensity rise
 in an achromatic fashion over  a period
$\delta t
\sim 3$ $\sqrt{M/.001 M_\odot}$ days.
Indeed, microlensing candidates have been found \cite{macho}. For low mass
objects, those with $M < 0.1M_\odot$, it appears however that 
the halo fraction
of MACHOs is small, $\approx 0.19^{+.16}_{-.10}$ \cite{m1}. 
Recent results from the MACHO collaboration \cite{newmacho}
show events with longer duration comprising about half the halo,
leading to the
speculation of a white dwarf population in the halo. 
Though it is too early to
determine the implications of these observations, 
they are very encouraging in
that perhaps this issue can and will be decided.

If we now take as given that non-baryonic dark matter is required, 
we are faced
with the problem of its identity.
	Light neutrinos ($m \le 30 eV$) are 
a long-time standard when it comes to
 non-baryonic dark matter \cite{ss}.  Light neutrinos produce 
structure on large scales, and the natural (minimal) scale for
 structure clustering is given in Eq. (\ref{mj}).  Hence neutrinos
 offer the natural possibility for large scale structures \cite{nu1,nu2} 
including filaments and voids.  It seemed, however, that neutrinos
 were ruled out because they tend to
 produce too much large scale structure \cite{nu3}.
  Because the smallest non-linear structures have mass scale $M_J$ and 
the typical galactic mass scale is $\simeq 10^{12} M_\odot$, galaxies must 
fragment out of the larger pancake-like objects.  The problem is 
that in such a scenario, galaxies form late \cite{nu2,nu4} 
 ($z \le 1$) whereas
 quasars and galaxies are seen out to redshifts $z \ga 4$. 
Recently, neutrinos are seeing somewhat of a revival 
in popularity in mixed dark matter models.

In the standard model, the absence of a right-handed neutrino state precludes
the existence of a neutrino mass.  By adding a right-handed 
state $\nu_R$, it is
possible to generate a Dirac mass for the neutrino,
 $m_\nu = h_\nu v/\sqrt{2}$,
as for the charged lepton masses, where $h_\nu$ is the neutrino Yukawa coupling
constant, and $v$ is the Higgs expectation value.  It is also possible to
generate a Majorana mass for the neutrino when in addition to the Dirac mass
term, $m_\nu \bar{\nu_R} \nu_L$, a term $M \nu_R \nu_R$ is included.
In what is known as the see-saw mechanism, the two mass eigenstates are given
by $m_{\nu_1} \sim m_\nu^2/M$ which is very light, and $m_{\nu_2}
 \sim M$ which is
heavy.  The state $\nu_1$ is our hot dark matter candidate as 
$\nu_2$ is in general
not stable.

	The cosmological constraint on the mass of a
 light neutrino is derived from the overall mass density of 
the Universe.  In general, the mass density of a light particle $\chi$ can be
expressed as
\beq
	\rho_\chi  = m_\chi Y_\chi n_\gamma  \le \rho_c  = 
1.06 \times 10^{-5} {h_o}^2 {\rm GeV/cm}^3	
\eeq
where $Y_\chi = n_\chi/n_\gamma$ is the density of $\chi$'s relative
 to the density of
 photons, for $\Omega {h_o}^2 < 1$.  For neutrinos $Y_\nu = 3/11$,
 and one finds \cite{cows}
\beq
	\sum_\nu (\frac{g_\nu}{2}) m_\nu  < 93 {\rm eV} (\Omega {h_o}^2)	
\label{ml}
\eeq
where the sum runs over neutrino flavors.  All particles with 
abundances $Y$ similar to neutrinos will have a mass limit 
given in Eq. (\ref{ml}).

 It was  possible that neutrinos (though not any of the 
known flavors) could have had large masses, $m_\nu > 1$ MeV.  In 
that case their abundance $Y$ is controlled by $\nu,{\bar \nu} $
 annihilations \cite{lw}, 
for example, $\nu {\bar \nu} \rightarrow f {\bar f}$ via Z exchange.  
When the annihilations 
freeze-out (the annihilation rate becomes slower than the expansion rate 
of the Universe), $Y$ becomes fixed.  Roughly, $Y \sim (m\sigma_A)^{-1}$
 and $\rho \sim {\sigma_A}^{-1}$
 where $\sigma_A$ is the annihilation cross-section.  For neutrinos, 
we expect $\sigma_A \sim {m_\nu}^2/{m_Z}^4$ so that $\rho_\nu 
\sim 1/{m_\nu}^2$  and
 we can derive a lower bound \cite{lw,ko,wso} on the neutrino mass,
 $m_\nu \ga 3-7$ GeV, depending on whether it is a Dirac 
or Majorana neutrino. Indeed, any particle with roughly a weak scale
cross-sections will tend to give an interesting 
value of $\Omega h^2 \sim 1$.

  Due primarily to the limits from LEP \cite{lep}, the heavy 
massive neutrino has become simply an example and is no longer a
 dark matter candidate. LEP excludes neutrinos (with standard weak
interactions) with masses $m_\nu \la 45$ GeV.  Lab constraints for 
Dirac neutrinos are available \cite{dir}, excluding neutrinos 
with masses between
10 GeV and 4.7 TeV. This is significant, since it precludes the possibility 
of neutrino dark matter based on an asymmetry between $\nu$ and ${\bar \nu}$ 
\cite{ho}. Majorana neutrinos are excluded as {\em dark matter}
since $\Omega_\nu {h_o}^2 < 0.001$ for $m_\nu > 45$ GeV and are thus
cosmologically  uninteresting.

If we return to our example of right-handed neutrinos, we saw
in the previous section on constraints from BBN that because
right-handed interactions are weaker than standard left-handed 
interactions, they decouple early and today are at a reduced 
temperature relative to $\nu_L$ (cf. eq. (\ref{decx})).
As such, for ${T_d}_R \gg 1$ MeV, $n_{\nu_R}/n_{\nu_L} = 
(T_{\nu_R}/T_{\nu_L})^3 \ll 1$. Thus the abundance of right-handed
neutrinos can be written as 
\beq
Y_{\nu_R} = {n_{\nu_R} \over n_\gamma} = ({3 \over 11})
({T_{\nu_R} \over T_{\nu_L}})^3 \ll {3 \over 11}
\eeq
In this case, the previous bound (\ref{ml}) on neutrino masses is weakened.
 For a suitably large scale for the right-handed interactions,
 right-handed neutrino masses may be as large as a few keV \cite{ot}.

Supersymmetric theories introduce several possible candidates. 
If R-parity, which distinguishes between ``normal" matter and the 
supersymmetric partners and can be defined in terms of baryon, lepton and
spin as $R = (-1)^{3B + L + 2S}$, is unbroken, there is at least one 
supersymmetric particle which must be stable.  I will assume R-parity
conservation, which is common in the minimal supersymmetric standard model
(MSSM). R-parity is generally
 assumed in order
to justify the absence of superpotential terms 
which can be responsible for rampid proton decay. 
 The stable 
particle (usually called the LSP) is most probably some 
linear combination of the $R=-1$ neutral fermions, the 
neutralinos \cite{ehnos}: the wino $\tilde W^3$, the partner of the
 3rd component of the $SU(2)_L$ gauge boson;
 the bino, $\tilde B$, the partner of the $U(1)_Y$ gauge boson;
 and the two neutral Higgsinos,  $\tilde H_1$ and $\tilde H_2$.
 Gluinos are expected to be heavier --$m_{\tilde g} = (\frac{\alpha_3}{\alpha})
 \sin^2 \theta_W M_2$, where $M_2$ is the supersymmetry breaking SU(2) gaugino
mass--  and they do not mix with the other states.
  The sneutrino \cite{snu} is also a possibility but has been
 excluded as a dark 
matter candidate by direct \cite{dir} searches, indirect \cite{indir}
 and accelerator\cite{lep} 
searches. 

There are relatively few parameters in the minimal model and 
the identity of the LSP is determined by three:
\begin{enumerate}
\item $M_2$;
if one assumes a common soft supersymmetry breaking gaugino mass  at 
the unification scale in the Lagrangian, 
 ${\cal L} \ni M_2 \tilde W \tilde W$. 
\item $\mu$;
 a Higgs  mixing mass, ${\cal L} \ni \mu \tilde H_1 \tilde H_2$
necessary to avoid 
phenomenologically unacceptable axions.
 \item $\tan \beta$;
 the ratio of Higgs expectation values, $\tan \beta = 
\frac{v_1}{v_2}$,  where $v_1 = \langle H_1 \rangle , v_2 = \langle H_2 \rangle$
 is also a free parameter but can be chosen 
to be positive without loss of generality. In this notation, it is $H_1$ 
which is responsible for up quark masses so that it will be natural to 
assume $\tan \beta > 1$.
\end{enumerate}

	The LSP can  be expressed as a linear combination
\begin{equation}
		\chi = \alpha \tilde W^3 + \beta \tilde B + \gamma \tilde H_1 + \delta \tilde H_2
\end{equation}
Pure state LSP possibilities are : The photino\cite{phot}
\begin{equation}
		\tilde {\gamma} = \tilde W^3\sin\theta_W + \tilde B \cos\theta_W
\end {equation}
the Higgsino, $\tilde S^0$ \cite{ehnos}
\begin{equation}
		\tilde S^0 = \tilde H_1 \cos\beta + \tilde H_2 \sin\beta
\end{equation}
the bino\cite{osi3},
\begin{equation}
	          \tilde B = \tilde B
\end{equation}
and the Higgsinos $\tilde H_{12}$\cite{osi3}
\begin{equation}
		\tilde H_{12} = \frac{1}{\sqrt{2}}  (\tilde H_1 \pm \tilde H_2)
\end{equation}
	The solution for the coefficients $\alpha, \beta, \gamma$ and $\delta$
for neutralinos that make up the LSP 
can be found by diagonalizing the mass matrix
\beq
      ({\tilde W}^3, {\tilde B}, {{\tilde H}^0}_1,{{\tilde H}^0}_2 )
  \left( \begin{array}{cccc}
M_2 & 0 & {-g_2 v_1 \over \sqrt{2}} &  {g_2 v_2 \over \sqrt{2}} \\
0 & M_1 & {g_1 v_1 \over \sqrt{2}} & {-g_1 v_2 \over \sqrt{2}} \\
{-g_2 v_1 \over \sqrt{2}} & {g_1 v_1 \over \sqrt{2}} & 0 & -\mu \\
{g_2 v_2 \over \sqrt{2}} & {-g_1 v_2 \over \sqrt{2}} & -\mu & 0 
\end{array} \right) \left( \begin{array}{c} {\tilde W}^3 \\
{\tilde B} \\ {{\tilde H}^0}_1 \\ {{\tilde H}^0}_2 \end{array} \right)
\eeq
where $M_1$ is a soft supersymmetry breaking
 term giving mass to the U(1)  gaugino.
  In a unified
 theory $M_1 = M_2$ at the unification scale translates to
\beq
	M_1 = {5 \over 3}  {\alpha_1 \over \alpha_2}  M_2	
\eeq
at low energies.  By performing a change of basis
 to ${\tilde W}^3 ,{\tilde B}, {\tilde A}^0 ,{\tilde S}^0$  with
\beq
	{\tilde A}^0 = {{{\tilde H}^0}_1 \sin \beta - {{\tilde H}^0}_2} \cos \beta
\eeq
  the mass matrix simplifies and becomes
\beq
      ({\tilde W}^3, {\tilde B}, {{\tilde A}^0}, {{\tilde S}^0} )
  \left( \begin{array}{cccc}
M_2 & 0 & {-g_2 v \over \sqrt{2}} &  0 \\
0 & M_1 & {g_1 v \over \sqrt{2}} & 0 \\
{-g_2 v \over \sqrt{2}} & {g_1 v \over \sqrt{2}} & {2 v_1 v_2 \over v^2} \mu &
{ {v_2}^2 - {v_1}^2 \over v^2} \mu \\
0 & 0 & { {v_2}^2 - {v_1}^2 \over v^2} \mu & {-2 v_1 v_2 \over v^2} \mu
\end{array} \right) \left( \begin{array}{c} {\tilde W}^3 \\
{\tilde B} \\ {{\tilde A}^0} \\ {{\tilde S}^0} \end{array} \right)
\eeq
where $v^2 = {v_1}^2 + {v_2}^2$ and can be solved analytically.
The lightest mass eigenstate is the LSP. 

 There are some limiting cases in which the LSP 
is nearly a pure state.  When $\mu  \rightarrow 0$, ${\tilde S}^0$ 
 is the LSP with
\beq
	m_{\tilde S}  \rightarrow  {2 v_1 v_2 \over v^2} \mu
\eeq
When $M_2 \rightarrow 0$, the photino is the LSP with
\beq
	m_{\tilde \gamma} \rightarrow {8 \over 3}
{ {g_1}^2 \over  ( {g_1}^2 + {g_2}^2) } M_2
\eeq
When $M_2$  is large and $M_2 \ll \mu$  then the bino ${\tilde B}$
 is the LSP    and
\beq
	m_{\tilde B}  \simeq M_1 	
\eeq
and finally when $\mu$ is large and $\mu  \ll M_2$
 either the Higgsino state
\beq
	{\tilde H}_{(12)}  =  {1 \over \sqrt{2}}  ( {{\tilde H}_1}^0 +  
{{\tilde H}_2}^0)  \qquad         
 \mu < 0	
\eeq
or the state
\beq
{\tilde H}_{[12]}  =  {1 \over \sqrt{2}}  ( {{\tilde H}_1}^0 - 
{{\tilde H}_2}^0)  \qquad         
 \mu > 0		
\eeq
is the LSP depending on the sign of $\mu$.

In Figure 24 \cite{osi3}, regions in
the $M_2, \mu$  plane with $\tan\beta = 2$ are shown in which the LSP
is one of several nearly pure states, the photino, $\tilde \gamma$, the
bino,
$\tilde B$, a symmetric combination of the Higgsinos, 
$\tilde{H}_{(12)}$, or the Higgsino, 
$\tilde{S}$. The dashed lines show the LSP mass contours.
 The cross hatched regions correspond to parameters giving
  a chargino ($\tilde W^{\pm}, \tilde H^{\pm}$) state 
with mass $m_{\tilde \chi} \leq 45 GeV$ and as such are 
excluded by LEP\cite{lep2}.
This constraint has been extended by LEP1.5, \cite{lep15,efos} 
and LEP2 \cite{asusy,efos2} and is shown by the 
light shaded region and corresponds to regions where the chargino mass is $\la
80$ GeV. The newer limit does not extend deep into the Higgsino region
because of the degeneracy between the chargino and neutralino.
The dark shaded region corresponds to an older limit on
$M_2$ from the limit\cite{cdf}
 on the gluino mass $m_{\tilde g} \leq 70$ GeV or $M_2 \leq 22$ GeV.
 Notice that the parameter space is dominated by the  
$\tilde B$ or $\tilde H_{12}$
 pure states and that the photino (often quoted as the LSP)
 only occupies a small fraction of the parameter space,
 as does the Higgsino combination $\tilde S^0$.


\begin{figure}
\hspace{0.5truecm}
\epsfysize=11truecm\epsfbox{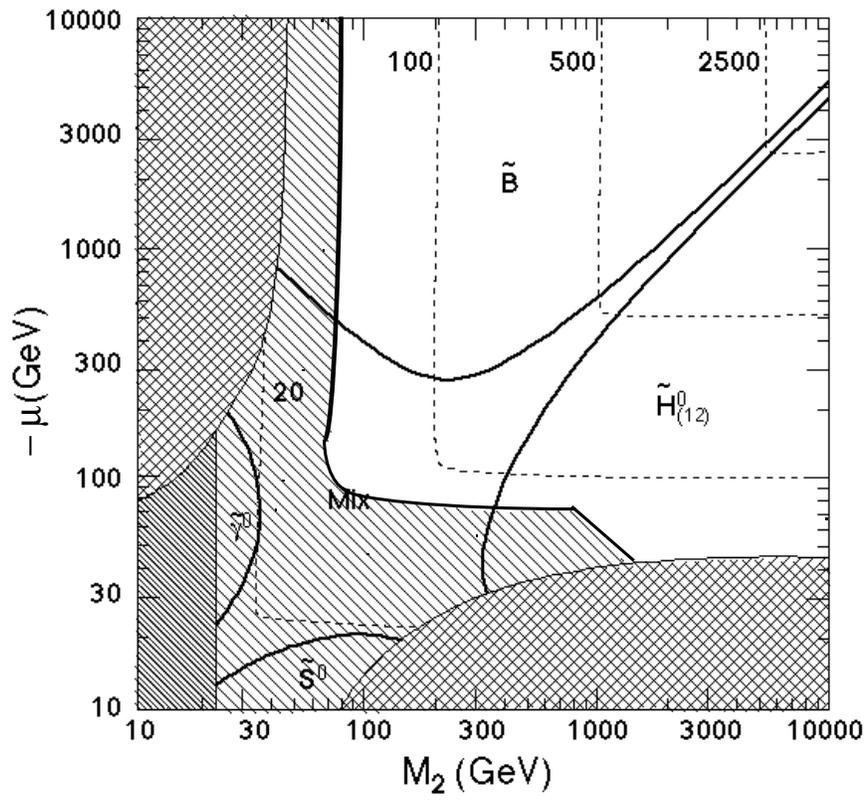}
\baselineskip=2ex
\caption { The $M_2$-$\mu$ plane in the MSSM for $\tan \beta = 2$.
}
\label{fig:fig16}
\end{figure}

\baselineskip=3ex

The relic abundance of LSP's is 
determined by solving
the Boltzmann
 equation for the LSP number density in an expanding Universe.
 The technique\cite{wso} used is similar to that for computing
 the relic abundance of massive neutrinos\cite{lw}.
The relic density depends on additional parameters in the MSSM.
These include the sfermion masses, $m_{\tilde f}$, the Higgs pseudo-scalar
mass, $m_A$, and the tri-linear masses $A$ as well as two phases
$\theta_\mu$ and $\theta_A$.
 For binos, as was the case for photinos \cite{phot}, it is possible
 to adjust the sfermion masses  to obtain closure density.
Adjusting the sfermion mixing parameters \cite{fkmos} or CP violating phases
\cite{fkos,fko} allows even greater freedom.
 In Figure 25 \cite{70}, the relic abundance ($\Omega h^2$) is shown in the
$M_2-\mu$
 plane with $\tan\beta = 2$, $m_{\tilde f} = 200$ GeV.
 Clearly the MSSM offers sufficient room to solve the dark matter problem.
Similar results have been found by other groups \cite{gkt,dn,dvn}.
In Figure 25, in the higgsino sector ${\tilde H}_{12}$ marked off by the dashed
line, co-annihilations \cite{gs,dn}
 between ${\tilde H}_{(12)}$ and the next lightest 
neutralino (also a Higgsino)
were not included. These tend to lower significantly 
the relic abundance in much
of this sector.

\begin{figure}
\hspace{0.5truecm}
\epsfysize=11truecm\epsfbox{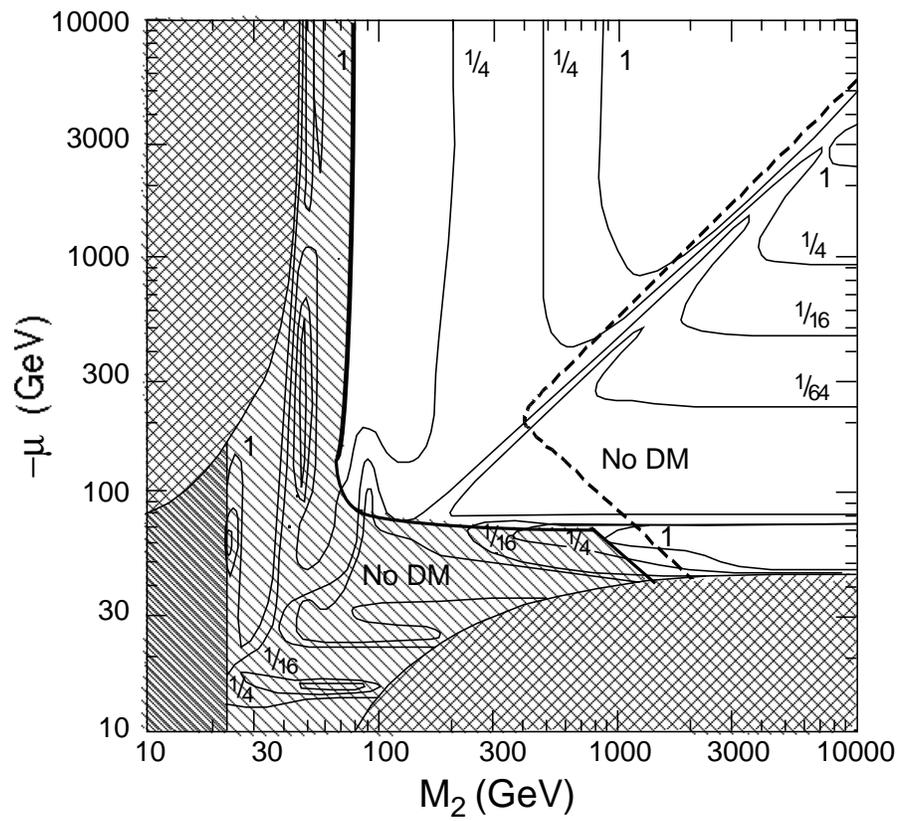}
\baselineskip=2ex
\caption { Relic density contours ($\Omega h^2$) in the $M_2$ - $\mu$ plane.
}
\label{fig:fig17}
\end{figure}

\baselineskip=3ex

As should be clear from figures 24 and 25, binos are a good and likely
choice for dark matter in the MSSM.  For fixed $m_{\tilde f}$,
$\Omega \ga 0.05$, for all $m_{\tilde B} = 20 - 250$ GeV
largely independent of $\tan \beta$ and the sign of $\mu$.
In addition, the requirement that $m_{\tilde f} > m_{\tilde B}$
translates into an upper bound of about 250 GeV on the bino mass
\cite{osi3,gkt}.  By further adjusting the trilinear $A$ and accounting for
sfermion mixing this upper bound can be relaxed \cite{fkmos}.
 For fixed $\Omega h^2 = 1/4$, we would require
sfermion masses of order 120 -- 250 GeV for binos with masses in the range
20 -- 250 GeV.  The Higgsino relic density on the other hand is largely 
independent of $m_{\tilde f}$.  For large $\mu$, annihilations into $W$
and $Z$ pairs dominate, while for lower $\mu$, it is the annihilations
into Higgs scalars which dominate.  Aside from a narrow region with 
$m_{\tilde H_{12}} < m_W$ and very massive Higgsinos with 
$m_{\tilde H_{12}} <250$ GeV, the relic density of ${\tilde H_{12}}$
is very low. Above about 1 TeV, these Higgsinos are also excluded. 

One can make a further reduction in the number of parameters by setting
all of the soft scalar masses equal at the GUT scale. In particular,
setting the Higgs soft masses equal to a common sfermion mass $m_o$
and enforcing the condition of radiative electroweak symmetry breaking
fixes $\mu$ and $m_A$ \cite{ir}. In this case, which we can refer to as the
constrained MSSM (CMSSM) the LSP is almost always a bino \cite{susygut}.
For a given value of $\tan \beta$, the parameter space is best described
in terms of the common gaugino masses $m_{1/2}$ and scalar masses $m_o$
at the GUT scale.  In Figure 26 (taken from ref. 111)
this parameter space is shown for $\tan
\beta = 2$.  The contours labeled by a value of $p$ represent
bino purity contours. The other contours are labeled by the value of
$\Omega h^2$. The shaded region corresponds to regions where either the
LSP is not a neutralino (but rather a stau) or there is a light chargino.

\begin{figure}
\hspace{0.5truecm}
\epsfysize=15truecm\epsfbox{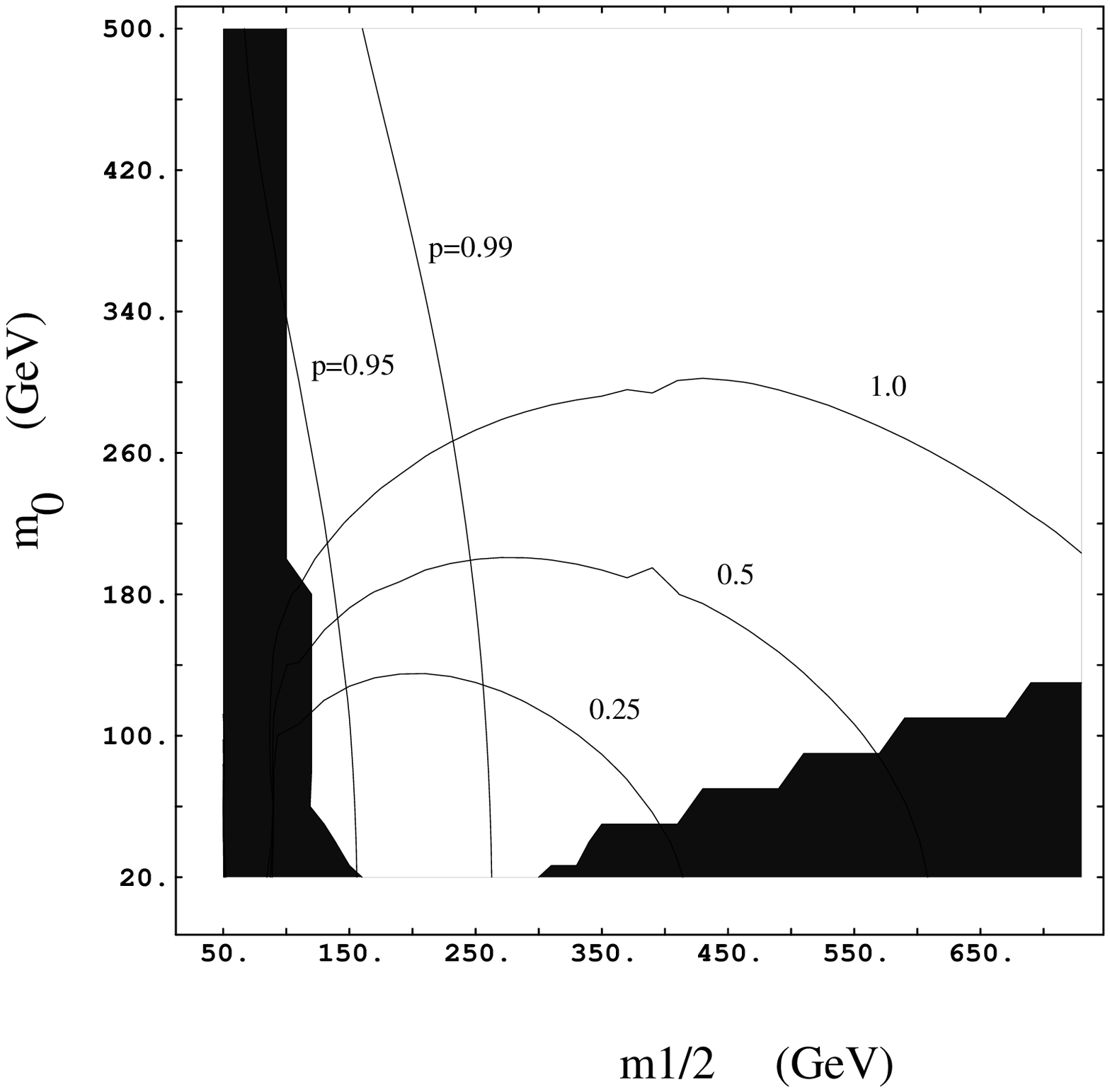}
\baselineskip=2ex
\caption { The $m_{1/2}$-$m_o$ plane in the CMSSM for $\tan \beta = 2$.
}
\end{figure}

\baselineskip=3ex

The $m_{1/2} - m_o$ parameter space is further constrained by the recent 
runs at LEP. In Figure 27, the constraints imposed by the unsuccessful
ALEPH chargino and neutralino
searches~\cite{achi} (long-dashed line) at LEP 1.5
are shown. This search could not
exclude regions of parameter space for small $m_o$, when the sneutrino mass
drops below the chargino mass.  However, some of this area
can be further excluded \cite{efos} by the limits  on
  $m_{\tilde \nu}$ (short-dashed line), by the LEP limits~\cite{asusy,lsusy}
  on slepton production (solid line), by 
single-photon measurements~\cite{amy}
  (grey line), and by the D0 limit on the 
gluino mass~\cite{d0} (dotted line). 
The region of the plane in which $0.1 < \Omega_{\chi} h^2< 0.3$ for some
  experimentally-allowed value of $\mu<0$ is light-shaded, and the region of
  the plane in which $0.1 < \Omega_{\chi} h^2< 0.3$ for $\mu$ determined by
  the CMSSM constraint on the scalar masses is shown dark-shaded.
  The constraint derived from the ALEPH searches when imposing CMSSM
constraints (labeled EWSB) is also shown as a solid line.
The improvement in these bounds \cite{efos2}
 by LEP 1.7 is illustrated in Figure 28. 

\begin{figure}
\hspace{0.5truecm}
\epsfysize=15truecm\epsfbox{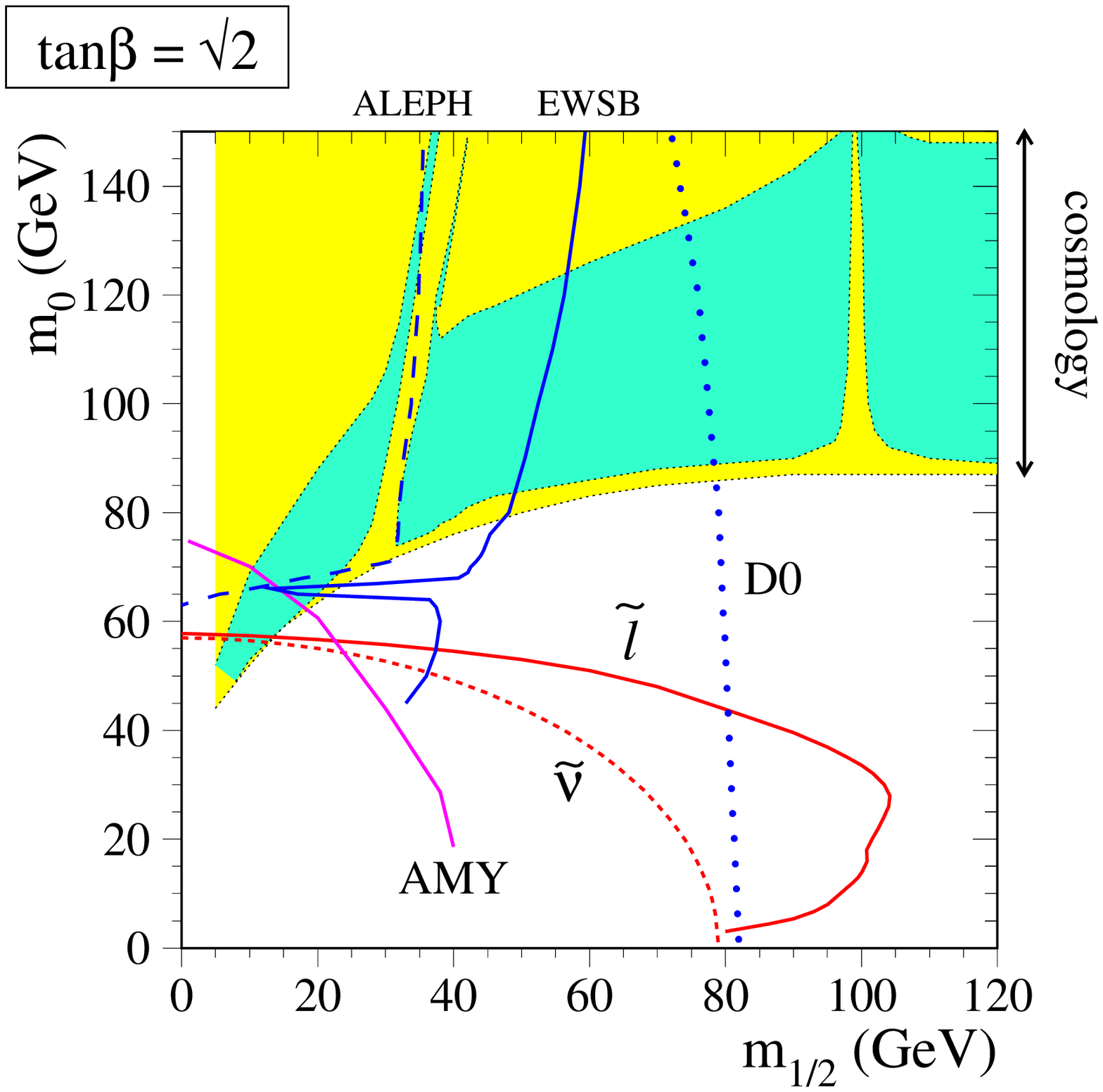}
\baselineskip=2ex
\caption { The $m_{1/2}$-$m_o$ plane
with constraints from Lep 1.5.
}
\end{figure}

\begin{figure}
\hspace{0.5truecm}
\epsfysize=11truecm\epsfbox{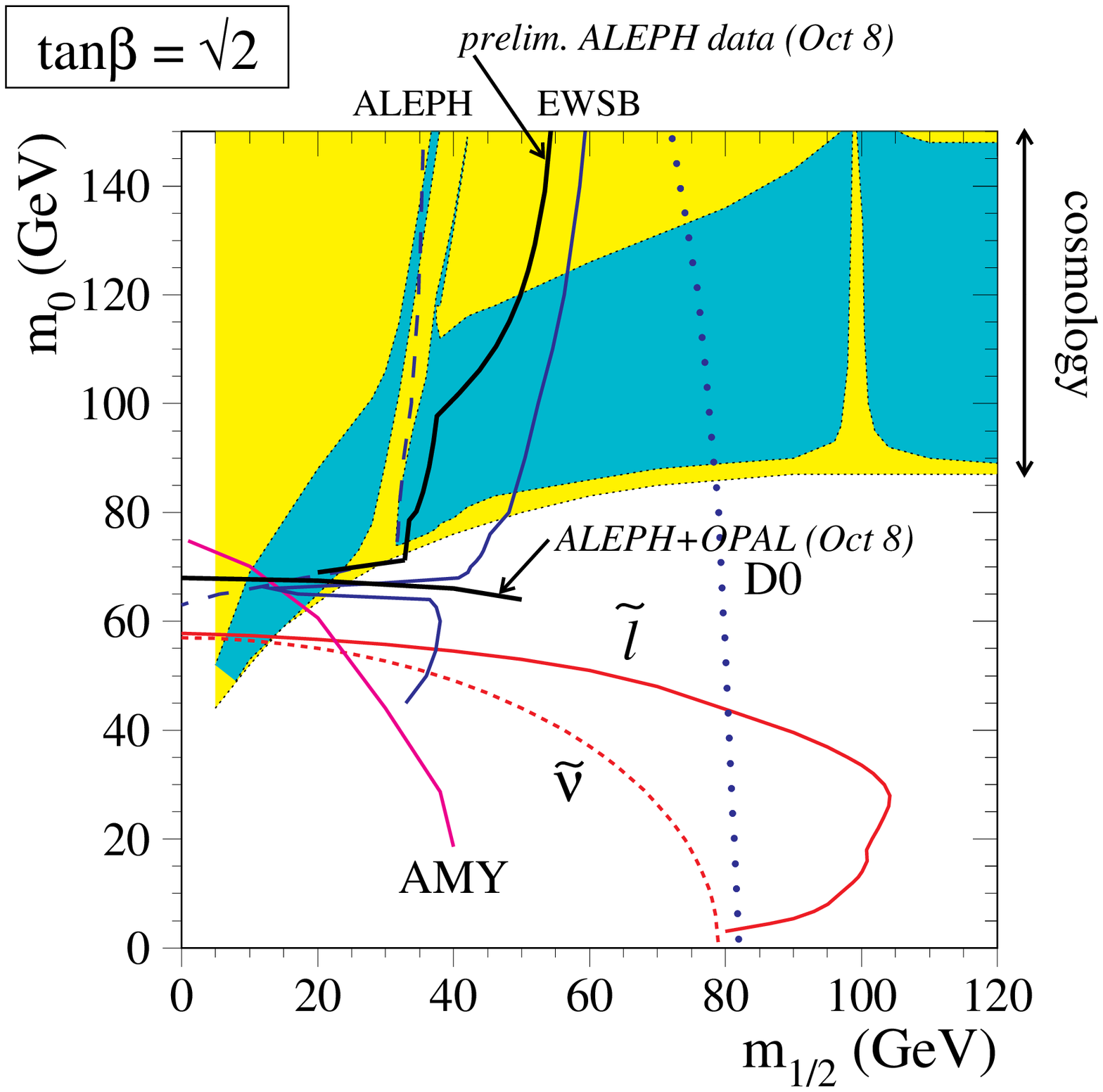}
\baselineskip=2ex
\caption { Same as Figure 27, but updated by the LEP 1.7 run.
}
\end{figure}

The final subject that I will cover in this lecture
is the question
of dark matter detection and this topic will be covered in 
detail by J. Ellis \cite{jellis}. 
The best chance for the detection of baryonic dark matter, as was discussed
earlier, is through gravitational microlensing \cite{pac,m1,newmacho}.
In principle, the duration of the signal should be able to determine
the mass of the macho and perhaps lead to a determination of the 
composition of the galactic halo.
The first test for the presence of neutrino dark matter is the experimental
verification of neutrino masses.  This will most likely be done first
by observing neutrino oscillations for the case of the light 
neutrinos (hot dark matter).  The heavy neutrinos, as explained earlier,
are already excluded as dark matter experimentally.

Other non baryonic dark matter detection can be 
separated into two basic methods,
direct and indirect.  Direct detection relies on the
ability to detect the elastic scattering of a dark matter 
candidate off a nucleus
in a detector.  
The detection rate will depend on the density of dark matter in the solar
neighborhood, $\rho \sim 0.3$ GeV/cm$^3$, the velocity, $v \sim 300$ km/s, and
the elastic cross section, $\sigma$.  Spin-independent 
interactions are the most
promising for detection.  Dirac neutrinos have spin-independent 
interactions, but
as noted above, these have already been excluded as dark matter by direct
detection experiments \cite{dir}. In the MSSM, it is possible for 
the LSP to also
have spin independent interactions which are mediated by 
Higgs exchange.  These
scatterings are only important when the LSP is a mixed 
(gaugino/Higgsino) state
as in the central regions of Figures 24 and 25. Generally, 
these regions have low
values of $\Omega h^2$ (since the annihilation cross sections
 are also enhanced)
and the parameter space in which the elastic cross section and 
relic density are
large is rather limited.  Furthermore, a significant detection rate 
in this case
relies on a low mass for the Higgs scalar \cite{bfg,fos}.

More typical of the SUSY parameter space is a LSP with spin dependent
interactions.  Elastic scatterings are primarily spin dependent 
whenever the LSP
is mostly either gaugino or Higgsino. For Higgsino dark matter, 
Higgsinos with
scatterings mediated by $Z^0$ avoid the
$\tilde{H}_{(12)}$ regions of Figures 24 and 25, and as such are now largely
excluded (the $S^0$ region does grow at low $\tan \beta$ \cite{ehnos,osi3}).
Higgsino scatterings mediated by sfermion exchange depend on couplings
proportional to the light quark masses and will have cross sections which are
suppressed by $(m_p/m_W)^4$, where $m_p$ is the proton mass. These rates are
generally very low \cite{fos}.  Binos, on the other hand, will 
have elastic cross
sections which go as $m^2/{m_{\tilde f}}^4$, where $m$ is the
 reduced mass of the
bino and nucleus. These rates are typically higher (reaching up to almost 0.1
events per kg-day \cite{fos,ef,bg}).

Indirect methods also offer the possibility for the detection of dark matter.
Three methods for indirect detection are often discussed.
 1) $\gamma$-rays
from dark matter annihilations in the galactic halo are a possible signature
\cite{gam}. In the case of the MSSM, unless the mass of the LSP
 is larger than
$m_W$, the rates are probably too small to be detectable over background.
  2) Dark matter will be trapped gradually in the sun, and
annihilations within the sun will produce high energy neutrinos which may be
detected \cite{slkosi}; similarly, annihilations within the earth 
may provide a
detectable neutrino signal \cite{earth}. 
  This method holds considerable promise, 
as there will
be a number of very large neutrino detectors coming on line in the future.
Finally, 3) there is the possibility that halo annihilations 
into positrons and
antiprotons in sufficient numbers to distinguish them from 
cosmic-ray backgrounds \cite{gam}.


\section*{Acknowledgments} 
This work was supported in part by DOE grant
DE--FG02--94ER--40823.


\end{document}